\documentclass[useAMS,usenatbib]{mn2e}
\usepackage{graphicx}
\usepackage{epsfig}

\newcommand{\HI}{H\,{\sc i}}
\newcommand{\HII}{H\,{\sc ii}}
\newcommand{\SII}{[S\,{\sc ii}]}
\newcommand{\SIII}{[S\,{\sc iii}]}
\newcommand{\OIII}{[O\,{\sc iii}]}
\newcommand{\OII}{[O\,{\sc ii}]}
\newcommand{\OI}{[O\,{\sc i}]}

\newcommand{\NII}{[N\,{\sc ii}]}
\newcommand{\HeII}{He\,{\sc ii}}
\newcommand{\HeI}{He\,{\sc i}}
\newcommand{\NeIII}{[Ne\,{\sc iii}]}
\newcommand{\ArIII}{[Ar\,{\sc iii}]}
\newcommand{\ArV}{[Ar\,{\sc v}]}
\newcommand{\NeV}{[Ne\,{\sc v}]}
\def\p0{\phantom{0}}

\title[A New Population of Planetary Nebulae Discovered in the Large Magellanic Cloud (II) ]{A New Population of Planetary Nebulae Discovered in the Large
Magellanic Cloud (II): Complete PN Catalogue}
\author[Warren A. Reid and Quentin A. Parker]{Warren A. Reid$^{1}$\thanks{e-Mail:
warren@ics.mq.edu.au; tosame@bigpond.net.au } and Quentin A.
Parker $^{1,}$$^{2}$\thanks{e-Mail: qap@ics.mq.edu.au}\\
$^{1}$Department of Physics, Macquarie University, Sydney, NSW 2109, Australia\\
$^{2}$Anglo-Australian Observatory, PO Box 296, Epping, NSW 1710
Australia}
\begin{document}

\date{Accepted 2006 September 18. Received 2006 September 4; in original form 2006 July 4}

\pagerange{\pageref{0}--\pageref{0}} \pubyear{2006}

\maketitle

\label{firstpage}

\begin{abstract}

This paper presents accurate homogeneous positions, velocities and
other pertinent properties for 460 newly discovered and 169
previously known planetary nebulae (PNe) in the central 25deg$^{2}$
bar region of the Large Magellanic Cloud (LMC). Candidate emission
sources were discovered using a deep, high resolution H$\alpha$ map
of the LMC obtained by median stacking a dozen 2 hour H$\alpha$
exposures taken with the UK Schmidt Telescope (UKST). Our
spectroscopic followup of more than 2,000 compact (ie. $<$20 arcsec)
H$\alpha$ emission candidates uncovered has tripled the number of
PNe in this area. All of the 169 previously known PNe within this
region have also been independently recovered and included in this
paper to create a homogeneous data set. Of the newly discovered PNe,
we classify 291 as ``true", 54 as ``likely" and 115 as ``possible"
based on the strength of photometric and spectroscopic evidence.
Radial velocities have been measured using both weighted averaging
of emission lines and cross-correlation techniques against high
quality templates. Based on the median comparison of the two
systems, we define a measurement error of $\pm$4 km s$^{-1}$. A new
velocity map of the central 25 deg$^{2}$ of the LMC, based on
results from the combined new and previously known PNe, is
presented, indicating an averaged heliocentric velocity differential
of 65 kms$^{-1}$ perpendicular to the line of nodes for the entire
PN population across our survey area. Averaged velocities of our PNe
and molecular hydrogen (from the literature) across 37 $\times$ 37
arcmin sub areas are compared. The PNe are found to have a higher
vertical velocity dispersion than the \HI~disk to a maximum of 10
times the spread of the \HI~disk, in keeping with the findings of
Meatheringham et al. (1988). In addition, moving out from the main
bar, we find that the PNe population follows a plane which is
somewhat warped in relation to the \HI~disk. We estimate the total
PN population of the entire LMC system, based on our R$_{equiv}$
H$\alpha$ limiting magnitude of $\sim$22, to be 956$\pm$141.
\end{abstract}


\section{Introduction}

The Large Magellanic Cloud may hold the answers to many questions
regarding the physical properties of planetary nebulae (PNe) due to
it's known distance of 50kpc (Madore \& Freedman 1998, Mould et al.
2000) and low reddening environment (e.g. Kaler \& Jacoby, 1990).
The LMC is essentially a thin ($\sim$500pc) disk inclined at only
35~degrees to our line of sight (van der Marel \& Cioni 2001) so all
LMC PNe can be considered to reside at a similar distance. This
provides a single environment in which PN mass loss history (and
hence that of intermediate to low mass stars) can be studied in
detail in the context of both stellar and galactic evolution
(Jacoby, 2006).

The central 25deg$^{2}$ of the LMC galaxy is able to be imaged in
its entirety with a wide-field telescope such as the UK Schmidt
Telescope (UKST). Furthermore, it is sufficiently close to enable
PNe to be detected and resolved using current ground-based
telescopes. Prior to our recent work, the number of LMC PNe remained
modest ($\sim$300) and comprised a small fraction of the expected
total (estimated at $1000\pm250$, Jacoby 1980, $\sim$3000 Frew,
private communication). However, a large fraction of all LMC PNe in
the central bar region has now been discovered. With this large and
varied population, meaningful quantitative determinations of the PN
luminosity function, distribution, abundances, excitation ratios,
kinematics and crucially mass-loss history, can be estimated
precisely for the first time.

We used specially constructed deep, homogeneous, narrow-band
H$\alpha$ and matching broad-band `SR' (Short Red) maps of the
entire central 25 deg$^{2}$ of the LMC (for details, see Parker et
al. 2005). The survey used an exceptional quality, monolithic, 70\AA
~FWHM H$\alpha$ interference filter (Parker \& Bland-Hawthorn 1998)
and fine grained Tech-Pan film as detector (Parker and Malin 1999)
to yield maps with a powerful combination of sensitivity, resolution
and area coverage. These unique maps were obtained by co-adding
twelve well-matched UKST 2-hour H$\alpha$ exposures and six
15-minute equivalent SR-band exposures on the same field, taken over
a 3 year period. The `SuperCOSMOS' plate-measuring machine at the
Royal Observatory Edinburgh (Hambly et al. 2001) scanned, co-added
and pixel matched these exposures creating 10$\mu$m (0.67~arcsec)
pixel data which goes 1.35 and 1 magnitudes deeper than individual
exposures, achieving the full canonical Poissonian depth gain, e.g.
Bland-Hawthorn, Shopbell \& Malin (1993). This gives a depth
$\sim$21.5 for the SR images and $R_{equiv}\sim$22 for H$\alpha$
($4.5\times10^{-17}ergs~cm^{-2}~s^{-1}~$\AA$^{-1}$) which is at
least 1-magnitude deeper than the best wide-field narrow-band LMC
images previously available. The influence of variable stars was
alleviated and emulsion defects removed by median stacking exposures
taken over the 3 year period. An accurate world co-ordinate system
was applied by matching small point sources across an
astrometrically calibrated SuperCOSMOS sub-image to the same point
positions on the SR and H$\alpha$ UKST stacked images. This yielded
sub-arcsec astrometry, essential for success of the spectroscopic
follow-up observations.

Prior to these results there have been 7 kinematic studies of the
LMC PNe population which in total represent $\sim$138 objects. We
present new, high quality radial velocities for the 168 previously
known and 424 of the newly discovered PNe, thereby providing a 400\%
increase in LMC PN velocity data. Since LMC PNe are an object
population with an intermediate age between \HI~clouds with young
star clusters and the old population II clusters, the extra velocity
data is serving to answer questions regarding different LMC rotation
solutions and velocity gradients for different populations of
different ages (Freeman, Illingworth and Oemler (1983);
Meatheringham et al. (1988). Radial velocities from our large
numbers of new LMC PNe have enabled us to distinguish velocity
gradients for the young and old PNe populations, the results of
which may be applied to the kinematic movement of stellar
populations.

\subsection{Candidate Selection Technique}

The full 25 deg$^{2}$ area of our LMC deep map was subdivided into
16 separate, non overlapping image cells on a 4 x 4 grid, each with
$\textit{x}$ and $\textit{y}$ dimensions of approximately
1$^{\circ}$18$^{\prime}$. Candidate emission sources were found
within each cell using an adaptation of a technique available within
{\scriptsize KARMA} (Gooch 1996). The SR and H$\alpha$ narrow-band
greyscale {\scriptsize FITS} image maps were each assigned an
individual colour. We chose red for the SR map and blue for the
H$\alpha$ map. Careful selection of software parameters allowed the
intensity of each map to be perfectly balanced allowing only
peculiarities of one or other pass-band to be observed and measured.
Stars without strong emission lines, become a uniform, combined
colour while emission objects remain the assigned H$\alpha$ colour.
PNe and other compact emission objects gained a distinctive `halo'
defined by the colour assigned to the H$\alpha$ image. For more
details on the method, see Reid \& Parker (2006)(RP1 hereafter).

\label{section 1} 

\section[]{Spectroscopic Followup Observations}

Having identified more than 2,000 compact (dia. $\leqslant$20
arcsec) emission sources in the LMC, spectra were required to
confirm emission character and object types. A major spectral
confirmation program was undertaken in November and December 2004
comprising 5 nights using 2dF on the Anglo-Australian Telescope
(AAT), 7 nights using the 1.9m at South African Astronomical
Observatory (SAAO), 3 nights using the FLAMES multi-object
spectrograph on the ESO Very Large Telescope (VLT) UT2 and 7 nights
using the 2.3m Australian National University (ANU) telescope at the
Siding Spring Observatory (SSO). Finally, in February, 2005 we used
6dF on the UKST over three half nights as a final high dispersion
multi-object follow-up. For the purpose of this paper we are
presenting results obtained from the AAT and VLT with added
confirmation from the 1.9m SAAO and 2.3m SSO observations. In
addition, the VLT spectra, which probe faint emission lines in a
sub-sample of the newly identified PNe, plus the UKST 6dF high
resolution results, will be the subject of a further paper in this
series (Reid \& Parker in prep).

In Table~\ref{table 1} we briefly give details regarding the
spectroscopic follow-up observations, most of which are described in
this paper. The field names are observation identifications or
object names in the case of the 2.3m observations. The first three
2dF fields, named ST.., are service time runs and were also the
subject of RP1. Observations using 2dF on the AAT are named A
through to O and FLAMES observations on the VLT are named 1 to 9.
The fields for each of these multi-fibre observations have different
central coordinates. The FLAMES observations were centred on several
of the densest areas of the main bar.

\begin{table*}
\caption{Observing Logs for LMC Planetary Nebulae.}
\begin{tabular}{|l|c|c|c|c|c|c|c|c|c|c|}
  \hline \hline
   &   &   &    &   &    &   &    &   \\
  Field Name & Telesc. & Date & Grating  &  Dispersion &Central  &
   T$_{exp}$ & N$_{exp}$ & N$_{obj}$  \\
     &   &   &  Dispenser &  \AA/pixel & $\lambda$ (\AA)  & s  &
       &     \\
   \hline
  ST1 &  AAT & 26 Nov-03 & 300B & 4.299 & 5841 & 1500 & 2 & 131 \\
  ST2 &  AAT & 26 Nov-03 & 300B &  4.299 & 5841 & 1500 & 2 & 80 \\
  ST3 &  AAT  & 15 March-03 &  300B  &  4.299  &  5852  &  1800  &
  2  &  81  \\
     a1550,061-213      &   1.9m    &   09-13 Nov-04   &   300 &   5   &   5800    &   800 &   2   &   11  \\
a1550,214-324      &   1.9m    &   11-15 Nov-04   &   1200    &   1   &   6563    &   1000    &   2   &   10  \\
FLAMES 1-9   &  VLT   &  5-7 Dec-04 &  LR2   &  0.339  &  4272  &  1000    &   3  & 420  \\
FLAMES 1-9  &   VLT   &  5-7 Dec-04  & LR3   &  0.339  &  4797  &  1000    &   3  & 420  \\
FLAMES 1-9   &  VLT   &  5-7 Dec-04  & LR6   &  0.339  &  6822  &  1000    &   3 &  420  \\
2dF A-O   &     AAT &   13-16 Dec-04   &   300B    &   4.3    &   5852 &   1200    &   3   &   3603 \\
2dF-1200R A-O &     AAT &   17-18 Dec-04   &   1200R   &   1.105   &   6793.51 &   1200    &   2   &   3303 \\
RP   &   2.3m    &   07-18 Jan-05   &   600R+B  &   2.2 &   4600 + 6563 &   900 &   2   &   56   \\
6dF 1-3  &   UKST  &  3-5 Feb-05   &    425R    &   0.62    &    6750 &     600    &   3 &  573   \\
6dF 1-3  &   UKST  &  3-5 Feb-05   &    580V    &  0.62     &    4750 &     600   &   3 &  573   \\
    \hline
\end{tabular}
\label{table 1}
\end{table*}

\subsection{2dF Observations}

For the 2dF (Lewis et al. 2002) observations, all 2,000 selected
emission sources were plotted over the LMC in order that central
positions for the 2 degree diameter fields might be most efficiently
determined. An algorithm produced by the Anglo-Australian
Observatory (AAO) based on simulated annealing was also used to
establish the optimum tiling in order to minimise the number of 2dF
pointings and maximise the areal coverage. The simulation confirmed
that 15 2dF field set-ups would be required to cover all the objects
in the 25 deg$^{2}$ field. Due to the 2dF fields being circular
there was considerable overlap of observations resulting in many
objects within the central bar being observed up to three times.
These multiple observations of the same object from different fields
at different positions on the 2dF field plate using different fibres
were considered beneficial as an internal check on data integrity.

For each 2dF field position observed, 400 objects at a time
including targets, guide stars and dedicated sky fibres could be
allocated onto an interchangeable field plate and sent to 2
spectrographs, each of which images 200 fibres. An algorithm in the
2dF {\scriptsize CONFIGURE} program allocated the fibres to the
input positions, giving priority to objects as specified in the
input file. In several instances, candidate sources could not be
configured due to adjacent spacing constraints for 2dF fibre
buttons. After all 15 fields had been observed twice for combining,
three fields were re-observed twice in order to include candidates
which had been omitted due to crowding. All 21 2dF 300B fields
observed provided us with effectively $\sim$4,000 spectra. Each
field was observed 2 or 3 times for the purpose of combination.
Individual exposure times were mostly 1200s using the 300B grating
with a central wavelength of 5852\AA~and wavelength range
3600-8000\AA~ at a dispersion of 4.30\AA/pixel. The service time
spectra (ST1-ST3) had longer exposure times (1500s and 1800s). These
low-resolution observations, at 9.0\AA~FWHM, were used as the
primary means of object classification.

All 15 fields including the three repeated fields, were then
re-observed twice using the 1200R high resolution grating with a
central wavelength of 6793\AA. These observations cover a range
6200-7300\AA~with a dispersion of 1.10\AA/pixel and resolution of
2.2\AA~FWHM which cleanly separates the \SII 6716 and 6731 lines
used for electron density determination. The high resolution spectra
were also used for determination of systemic velocities as described
in section~\ref{section 4}. In all we had 7521 high and low
resolution object spectra from 2dF observations.

\subsection{Accuracy of Fibre Positioning}

The accuracy of fibre positioning on a target is dependent upon the
accuracy of the guide star positions, the central position for each
source to be observed and accuracy of the robotic arm which places
the fibres onto the field plate. Accurate astrometry for all objects
(to 0.67 arcsec) were obtained using our SuperCOSMOS digitised
H$\alpha$ map and the co-ordinate solution applied directly to it.
Guide stars at the required magnitudes for 2dF were chosen from the
ESO guide star catalogue. Their positions were double checked in
case adjustment was required to match the SuperCOSMOS digitised
H$\alpha$ map from which our candidate positions were determined to
ensure all sources (targets and fiducials) were on the same
astronomic grid. The 2dF robotic arm positions each fibre to an
accuracy of 0.3 arcsec and 2dF fibres have a diameter of $\sim$2
arcsec on the sky. This was advantageous as most fibres were able to
completely cover each discrete emission source, however several were
contaminated by superimposing stars within the field of view due to
the high stellar density of the LMC.

\subsection{Flames Observations on the VLT}

Spectroscopic observations of a sub-sample of 420 candidate PNe,
compact \HII~regions and WR stars in dense regions of the LMC main
bar were undertaken using FLAMES (Pasquini 2002) on the VLT UT2 over
three nights. The FLAMES multi-object spectrograph can observe
targets over a large corrected 25 armin dia field of view. Using the
GIRAFFE spectrograph, 132 targets at a time were configured onto a
fibre positioner, (OzPoz) hosting two rotating field plates. With
GIRAFFE, we used the MEDUSA fibre feeding system where each fibre
has an aperture of 1.2 arcsec on the sky. Each fibre allows a
minimum object separation of 10.5 arcsec determined by the size of
the magnetic buttons. This permitted us to gain individual spectra
from very close objects in a single observation.

The OzPoz is able to position the fibres with an accuracy of better
than 0.1 arcsec. With careful astrometric precision, therefore, we
were able to place the fibres on the precise position of the PNe or
PN halo that we desired. We used the low resolution (600 lines/mm)
grating on each field. The three filters used were LR2, LR3 and LR6
where each have a resolution of R=6400, 7500 and 8600 respectively.
Filters LR2 and LR3 allowed us to cover the most important optical
diagnostic lines for PNe in the blue including \OIII~4363\AA,
\HeII~4686\AA, H$\beta$ and \OIII~4959 \& 5007\AA. Filter LR6
covered the H$\alpha$, \NII~6548 and 6583\AA~lines as well as the
\SII~6716 \& 6731\AA~lines for electron densities. Other setup
details may be seen in table~\ref{table 1}.

\subsection{Other Observations}

Our first spectroscopic follow-up observations were conducted in
November 2004 using the 1.9-m telescope at SAAO. With a fixed E-W
slit and a low-dispersion grating, the 1.9-m provided 3800-7800\AA,
spectra for 23 compact and 43 extended sources (d$\leqslant$26
arcsec). Many of these objects were later observed on 2dF. Following
the December 2004 2dF run, there were several uncertainties raised
by obscuring stars in close proximity to candidate PNe which were
captured within the 2~arcsec diameter of the fibre. These problems
were solved by observing these candidates again using low resolution
observations on the  2.3-m Advanced Technology Telescope at SSO in
January 2005. The 2.3-m telescope with the double-beam spectrograph
(DBS) gave us the flexibility to visually orient the spectrograph
slit across the PN, missing any intervening stars. The visible range
(3200-9000\AA) of this spectrograph is split by a dichroic at around
6000\AA~and fed into two spectrographs, with red and blue optimized
detectors, respectively. We used the 300B (300lpmm) and 316R
(316lpmm) gratings to obtain a resolution of 5\AA~for each arm of
the DBS. Many candidate PN identifications were confirmed using
these telescopes but reliable radial velocity measurements require
higher dispersion data.

\label{section 2} 

\section[]{Data Reduction}

The 2dF raw data was processed using the AAO 2dF data reduction
system (2dFDR) found at
http://www.aao.gov.au/AAO/2df/software.html\#2dfdr. The automatic
reduction display loads an interactive dialog where files are loaded
and parameters set. On startup, the system creates the necessary
calibration groups. Each of these contain reduced calibration files
of different types (eg. BIAS, DARK, FLAT, ARC etc). Whenever a
calibration exposure is reduced it is inserted into the appropriate
group.

The heart of the reduction process is the generation of a tram-line
map which tracks the positions of the fibre spectra on the CCD. This
is generated using an optical model for the spectrograph and a file
listing the positions of the fibres on the slit. The fibre
extraction process then uses the tram-line map and the image to
extract the spectrum for each of the 200 fibres. Cosmic rays were
rejected during the process of combining two or more observations.

Reduction and extraction of 2.3m and 1.9m spectra were performed
using the standard {\scriptsize IRAF} tasks {\scriptsize IMRED,
SPECRED}, {\scriptsize CCDRED} and {\scriptsize FIGARO}'s task
{\scriptsize BCLEAN} through the PNDR\footnote[1]{The PNDR package,
including a manual and the code, can be found at:
http://www.aao.gov.au/local/www/brent/pndr/} package, an
{\scriptsize IRAF}-based PN reduction package, developed at
Macquarie University and the AAO, to handle the reduction of large
numbers of PN more efficiently. One dimensional spectra were created
and the background sky was subtracted. Final flux calibration used
the standard stars, LTT7987, LTT9239, LTT2415 and LTT9491.

The VLT FLAMES spectra was reduced using {\scriptsize IRAF} tasks
{\scriptsize IMRED, SPECRED} and {\scriptsize CCDRED} for multi-spec
files. Cosmic rays were rejected when combining frames. Using the
weighted intensity of the continuum, the {\scriptsize IRAF}
{\scriptsize SCOMBINE} task was then used to combine the three
different wavelength portions of the spectrum into one spectral
image.

\label{section 3}

\section[]{Results}

\begin{figure*}
  \includegraphics[width=1\textwidth]{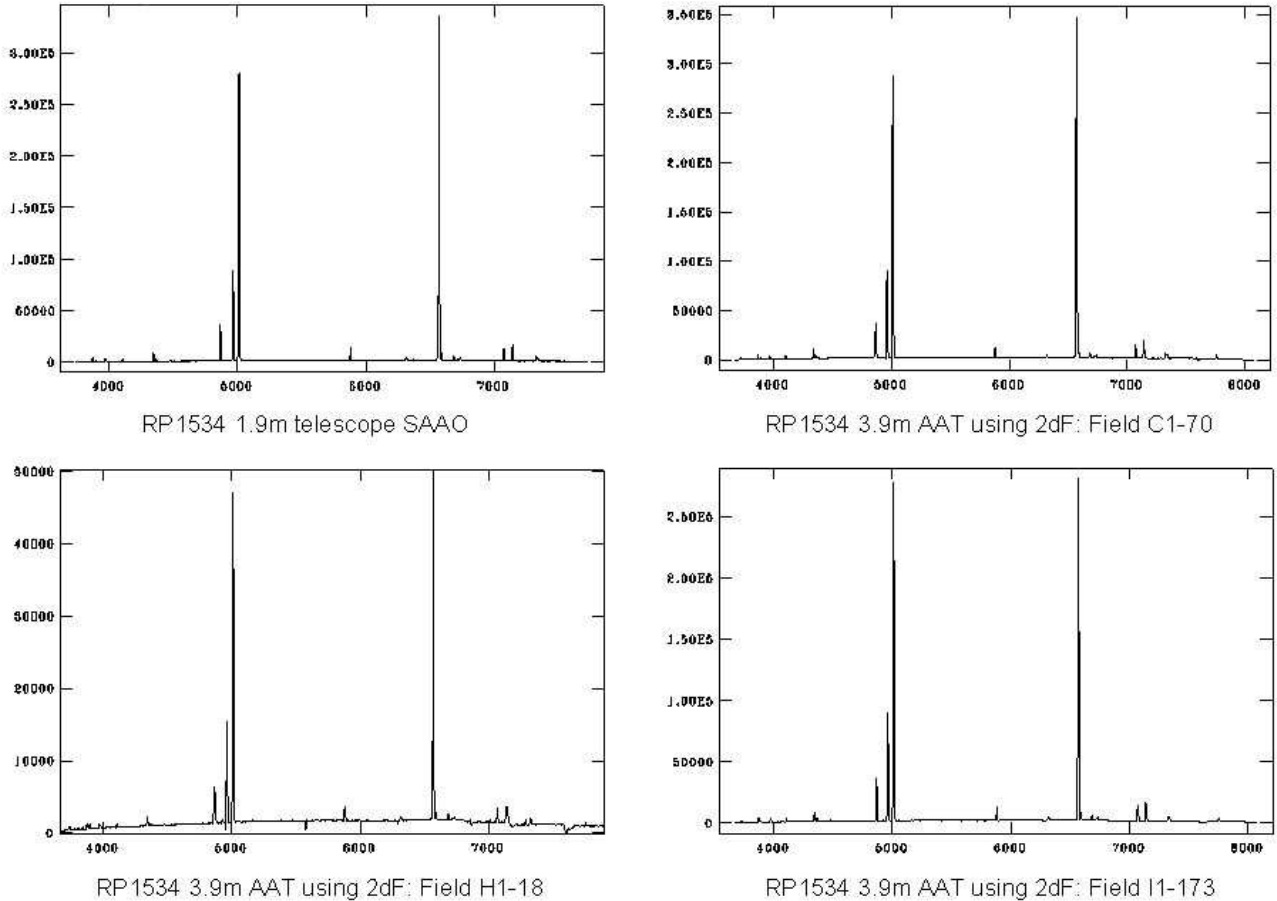}\\
  \caption{The comparison of four separate spectroscopic observations of RP1534, a newly discovered PN. One observation using the SAAO 1.9-m telescope is compared to three observations using 2dF on the AAT. The line strength ratios are consistent to within 97\%. }
  \label{Figure 1}
 \end{figure*}

Each candidate spectrum was examined carefully next to the matching
H$\alpha$ stacked image where we were able to account for anomalies
such as superimposed stars, seen in the spectrum. Where there were
two or more observations of an object, all the spectra were
nonetheless reduced and examined. These multiple spectra were used
to double check repeatability, consistency and velocities. Slight
changes in individual spectra from one observation to another were
mostly related to seeing conditions, where nearby stars would
encroach into the 2 arcsec fibre diameter. This was occasionally
noticeable in our multiple observations of the same object, where
emission line strengths (not ratios) or continuum from intervening
stars varied slightly in the spectral results. Comparison of 2.3-m,
1.9-m and 3.9-m data was generally excellent with emission-line
ratios consistent to within 97\%. Figure~\ref{Figure 1} shows one
example of a newly discovered PN, RP1534, which has been observed 4
times in low resolution. The spectrum on the upper left is from the
1.9-m SAAO telescope while the other three spectra are from
different 2dF field configurations on the AAT.

\subsection{The Confirmation of new PNe}

All 169 previously known PNe within the central 25deg$^{2}$ area
were also observed and measured for line intensities and velocities.
Of the 460 newly discovered PNe, we have classified 291 as ``true",
54 as ``likely" and 115 as ``possible" based on the following
analysis. To identify a planetary nebulae we first looked for higher
levels of ionization than are found in \HII~regions. We looked for
\HeII~plus strong collisionally excited lines such as \OIII,
\ArIII~and \NeIII~with strong \NeV~and \ArV~if present. We compared
intensities of the \OIII5007, \OIII4959 and H$\beta$ lines,
expecting a theoretical line intensity ratio of $\sim$9:3:1 for PNe.
This requirement was relaxed where \NII6583 $\geq$ H$\alpha$. In
these cases we also looked for the presence of
\HeII~4686\AA~indicating a possible high excitation Type I PN.
Diameters of the candidate sources were examined where we noted that
previously known LMC PNe have diameters up to 15 arcsec on the
stacked H$\alpha$ maps which represents an extent of $\sim$3.6pc at
the distance of the LMC. For bright, compact, stellar sources, much
of this image size could be attributed to the point spread function
as described in RP1. However, approximately 60\% of the PNe on our
H$\alpha$ map are surrounded by AGB halos and their radial extent
has also been included in our diameter estimates.

Candidate PNe that we considered to be ``true" satisfy all the
following criteria. They have a distinct, wide and dense H$\alpha$
emission shell compared to the size of the central source on the
merged H$\alpha$ and SR maps. Many ``true" PNe show no visible
central source at all when the H$\alpha$ and SR maps are merged
indicating very low stellar continuum levels on the SR map. The
H$\alpha$ emission also has a distinct boundary as opposed to the
appearance of emission-line stars where emission gradually fades
with increasing distance from the star. The measured diameters are
in close keeping with the mean diameter of the previously known PNe.
``True" PNe generally reside in areas where the intensity of the
H$\alpha$ emission from the PN is greater than 10 $\sigma$ any
averaged ambient diffuse emission. Spectroscopically, they have
either the required \OIII/H$\beta$ ratios or high \NII/H$\alpha$
ratios as described above. In the case of high \NII/H$\alpha$, a
strong \OIII~line has often been detected along with the high
excitation \HeII~4686\AA~line rarely seen in \HII~regions. The
\OII3727\AA~doublet was always detected while \NeIII3869\AA,
\ArIII7135\AA~and \HeI~6678\AA~were generally present and above 5
$\sigma$ (noise level). The \SII 6717 and 6731\AA~lines are clearly
present and measurable but not excessively high in comparison to
H$\alpha$. Line profiles were also examined using the 1200R grating
high resolution spectra. FWHM values for the strong emission lines
such as H$\alpha$ and \NII6583\AA~are $\leq$3\AA, corresponding
closely to measurements we made of previously known LMC PNe using
the same system.

Candidate PNe that we considered ``likely" were those with either
contaminating stars producing a stellar continuum in the spectrum or
those residing in areas of moderate-to-low \HII~density. In
appearance they are strong H$\alpha$ emitters with clear boundaries.
While most have strong \OIII/H$\beta$ ratios, in some cases the
ratio is down to 5:1. In a few cases, the \NII/H$\alpha$ ratio is
extremely high however \OIII 5007\AA~and H$\beta$ are relatively low
by comparison. Since many of these PN are Type I PN candidates,
internal extinction toward the blue end of the spectrum may be
partly the cause of the lower ratios. In addition, the destruction
of O during the ON cycle has been reported in massive progenitor
stars where it is strongly dependent on metallicity (Leisy et al.
2000). All ``likely" candidates have the \ArIII7135\AA~line which
traces the progenitor's metallicity as well as the
\OII3727\AA~doublet. Many also have \NeIII3869\AA, \HeI~6678\AA~and
a few have very low levels of \HeII~4686\AA.

Candidate PNe that we considered ``possible" are those either more
than 50\% obscured by stars, have \OIII5007\AA~levels between 3 and
5 times H$\beta$ or exist in areas of moderate-to-high \HII~density.
In appearance, many candidates in this category are extremely faint
and/or small. In this case, we may be exploring the faintest remains
of previously bright PNe. There are a wide range of spectral line
intensity ratios for this category, however lines of
\NeIII3869\AA~and \HeI~6678\AA~are usually $<$2 $\sigma$. Lines of
H$\alpha$, \OIII4959, 5007\AA~and \OII3727\AA~remain present in
acceptable ratios along with the \SII6717, 6731\AA~doublet in most
cases. We have also included a sub-class of ``possible" PNe called
very low excitation (VLE) PNe. These objects share most of the
dimensional and morphological characteristics of both known and new
PNe that we have observed on the stacked and merged maps.
Spectroscopically however, they are very low in \OIII4959 \&
5007\AA. The \OIII5007\AA~line must be present to be counted as a
VLE PN. In 8 of these objects, we found [OIII]5007\AA$>$H$\beta$ by
only $\sim$3 times the H$\beta$ line intensity. In 20 more suspect
objects we found \OIII5007\AA$\simeq$H$\beta$ line intensity. There
are 4 objects however, that we classified as VLE PNe, where
H$\beta$$>$\OIII5007\AA~by a factor of 2 times the \OIII~line
intensity. These objects are quite possibly extremely compact,
low-excitation \HII~regions, however their size and morphology are
more suggestive of PNe. None of them have strong ionizing central
stars so they are very weak in H$\alpha$ and two of them are visible
in H$\alpha$ only. These 4 objects are always referred to as
``VLE-H$\beta$$>$\OIII" throughout our database.

\subsection{The Identification of Other Emission Objects}

Among emission candidates we have found a large number of other
astrophysical objects apart from PNe, many of which had the
appearance of previously known and newly discovered PNe on the
merged H$\alpha$ and short red maps. Spectral confirmation has
revealed a large number of them to be compact \HII~regions,
emission-line stars, late-type (M) stars, compact and dense ejecta
from supernova remnants (SNRs), Wolf-Rayet shells, and outflow
emission from symbiotic stars. Although we observed as many newly
discovered emission objects as possible, most \HII~regions were
pre-determined as such and rejected as PNe on the basis of size and
morphology. Most of them are both larger than 25 arcsec on the
stacked H$\alpha$ map and irregular in shape. Principally, they
exhibit dust and dark nebulae lanes, stratified regions and
bow-shocks with (multiple) stellar content. Spectroscopically, they
have \OIII5007/H$\beta$ ratios greater than 6, so without reference
to the photometric data, could be mistaken for a PN. Other isolated
and compact \HII~regions were difficult to eliminate as PNe by
visual examination of the H$\alpha$ images alone. Only spectroscopic
confirmation was able to reveal them due to their \NII/H$\alpha$
ratio less than 0.7 (Kennicutt et al. 2000) and \OIII5007/H$\beta$
ratio less than 1.0. Unlike their larger counterparts, we find that
the compact \HII~regions $<$20 arcsec dia have \OIII5007/H$\beta$
ratios $<$3.

Wolf-Rayet (WR) stars occasionally appear as strong H$\alpha$
emitters on the UKST stacked H$\alpha$ map. They have a bright
ionising star which appears considerably larger than PNe on the
contemporous short red map. The wind-blown shells or processed
ejecta of these stars create a unique spectrum which may have line
ratios similar to PNe but on a continuum. The line profiles however
are particularly wide and \HeII~4686\AA~is often stronger than
\OIII5007\AA. Many previously known WR stars were identified and
several new ones were found (see Table~\ref{table 2}).

SNR candidates in the LMC were initially identified by their
morphology on the H$\alpha$ map. Their large scale size excluded
them from consideration as PNe however certain concentrated ``lumps"
of isolated emission in the vicinity of the large shells were
checked spectroscopically as they often had the appearance of
previously known PNe on the H$\alpha$ map. While 9 previously known
SNRs were found and observed, 18 new probable SNRs were also
identified by their spectral signatures. They appear to be part of
larger circular structures, often with filamentary morphology,
bow-shocks and clumped areas. Positional coincidence with
non-thermal radio sources and X-ray identification was confirmed in
each case. With the detection of strong \SII~relative to H$\alpha$,
OII~and OI~emission lines in our spectroscopic follow-up, they
satisfy the diagnostic line ratio criteria for an SNR as established
by Fesen et al. (1985).  With these results, combined with a large
physical extent, we have classified them as probable SNR. These
objects will be the subject of a separate publication.

Other nebulae which have a compact ($<$20 arcsec dia.) appearance on
the stacked H$\alpha$ map but do not fit any of the above
spectroscopic criteria were retained as ``emission objects of
unknown nature" in keeping with the original identification by
Henize (1956). From our survey, 23 of our objects were already
identified as ``emission objects of unknown nature" in the
{\scriptsize SIMBAD} online database. We have identified most of
these as \HII~regions. To the remaining 12 we add a further 25
objects. They may be grouped as follows: 1) those with the
appearance of a star but no continuum; 2) faint \HII~regions greater
than 27 arcsec diameter with very strong \OIII/H$\beta$; 3) small
($<$5 arcsec) and faint \HII~regions without \OII~or \OIII~lines.

A summary of the total number of objects in each classification is
shown in Table~\ref{table 2}. We have used our PN identification
criteria on every object spectrum including the previously known
PNe. Following multiple spectral observations, it is our opinion
that two objects be removed as PNe. The objects, indicated by the
(-2) in Table~\ref{table 2} are LM2-39, a late-type star with 5
observations confirming H$\beta$/\OIII5007 = 2.9 and SMP26 with 3
observations confirming H$\beta$/\OIII5007 = 72. We would also
classify previously known PN Sa114 with a strong continuum as
`likely'. A further 4 PNe; MG39 H$\beta$/\OIII5007 = 0.68 (6
observations);~SMP64 H$\beta$/\OIII5007 = 3.1 (5 observations);~
SMP77 H$\beta$/\OIII5007 = 0.82 (5 observations)~and SMP31 with
H$\beta$/\OIII5007 = 0.8 we categorize as `possible' PNe. The
largest group of objects found were emission line stars such as
CV's, T Tauri's, fo's and Be stars. These occasionally mimic the
appearance of PNe in the merged H$\alpha$ and SR maps. Most of them
however lack the well defined outer H$\alpha$ boundary observed
surrounding confirmed PNe. Due to the number of fibres available on
2dF, most of them were also observed. The spectroscopic observations
confirmed most of our estimated probabilities with regard to the
classification of emission line stars.

\begin{table}

\caption{Emission object classification results from spectral
observations covering the central 25~deg$^{2}$, area of the main LMC
bar as seen in Fig.~\ref{Figure 2}. By the application of our same
PNe analysis strategy as stated in section~\ref{section 3}, to the
previously known PNe, we would reduce the number by 2 objects,
classify 2 more as `Likely' 4 more as `Possible'.}

\begin{center}
\begin{tabular}{|l|c|cl}
  \hline\hline
  Object &  Previously &  Newly \\
  & Known &  Confirmed  \\
  \hline
  PNe `True' & 162(-2) & 291 \\
  PNe `Likely' & 1 & 54 \\
  PNe `Possible' & 4 & 115 \\
  Emission-line stars & 55 & 622\\
  Wolf-Rayet stars & 14 & 8 \\
  Late-type stars  & 10 & 247  \\
  Variable stars  & 61  & 28  \\
  Other stars   & 1 &  72  \\
  \HII~regions  & 85 & 69  \\
  Emission objects of &   & \\
  unknown nature & 12 & 25  \\
  SNR     &  9 &  18 \\
  S/N too low for ID  & &  32  \\
  \hline \noalign{\smallskip} 
  \label{table 2}
  \end{tabular}
\end{center}
\end{table}

\subsection{PN Population Estimates}

We have estimated the total number of PNe in the LMC using the ratio
of new to previously known PNe within the central 25 deg$^{2}$
region surveyed. There are 169 previously known and 460 new PNe
within this region. Since $\sim$300 were previously known in the
whole LMC system, $\sim$130 lie outside our survey region. All
together there are now 760 known and possible PNe in the LMC. If the
same ratio of discovery is possible outside our current survey area,
based on the same detection limit as the H$\alpha$ stack (m=22),
there may be as many as 1113 PNe in the LMC. The 169 PNe we have
listed as likely and possible, however, may be used as an error
estimate. The number of PNe within the survey region is then
541$\pm$89 and the number outside is 415$\pm$68. Together our
magnitude limited estimate is 956$\pm$141 for the entire LMC system.

\label{section 4}

\section{Main Bar PN Distribution}

\begin{figure*}
  \includegraphics[width=1\textwidth]{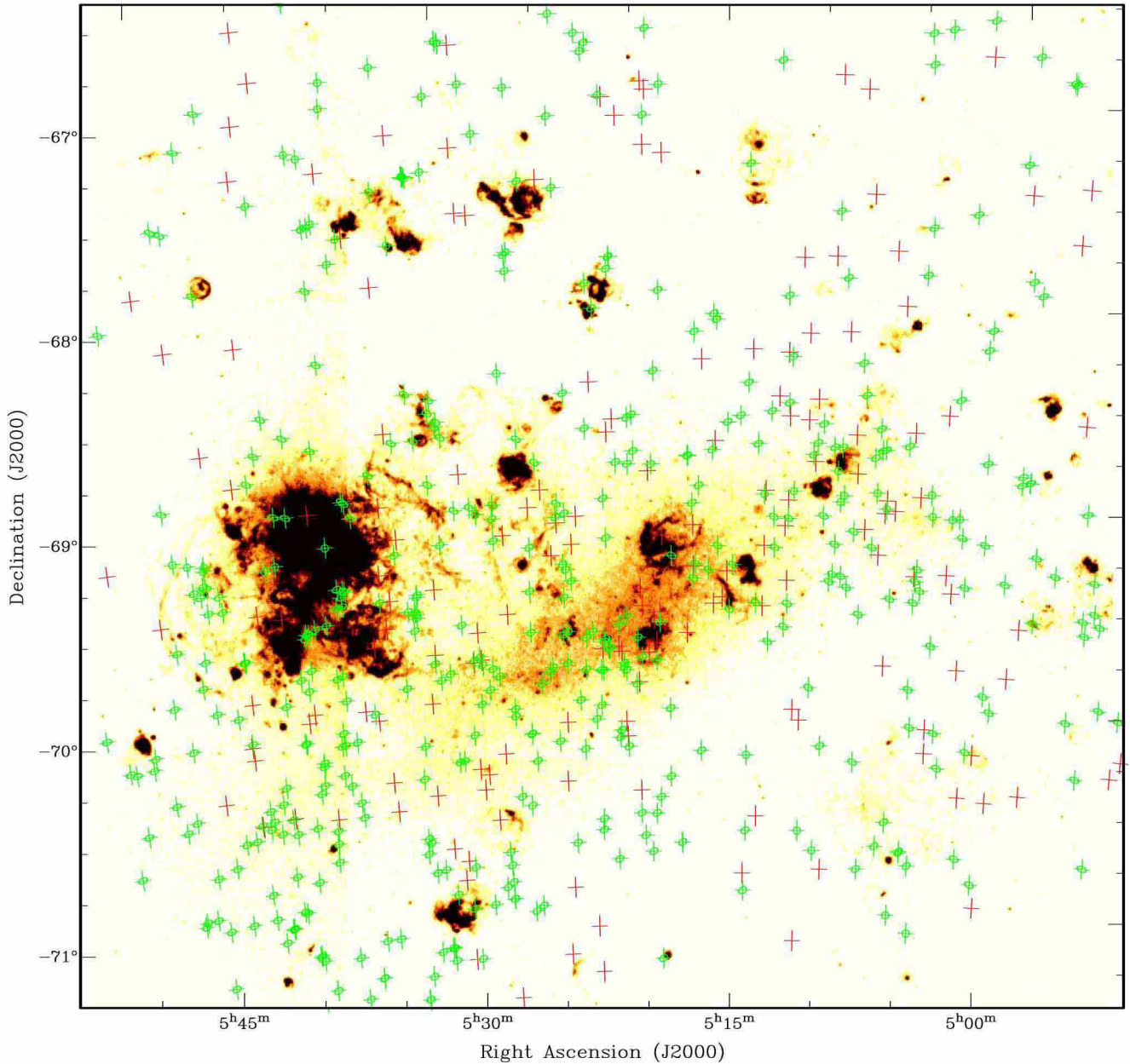}\\
  \caption{Previously known PNe (red crosses) and newly discovered PNe (green crosses with a $\oplus$) are plotted on the UKST H$\alpha$ map of the central 25 deg$^{2}$ of the LMC. The number density increases to the SW of the main bar axis where the plane of the PNe population is less inclined to our line of sight. Above the main bar region there is a NW-SE diagonal region in which the density of PNe is very low. This area corresponds with a steep gradient in velocity for both the PNe population and the \HI~gas.}
  \label{Figure 2}
 \end{figure*}

The distribution of PNe across our survey area is shown on the
H$\alpha$ map of the LMC in figure~\ref{Figure 2}. Red crosses mark
the positions of previously known PNe while the green circles
approximate the positions of our newly discovered PNe. The
distribution of newly discovered PNe somewhat follows the same
distribution seen with previously known PNe.  The increased density
of PNe toward the main bar is clear however this density extends in
a south-east direction below the 30Dor region where there are
extensive, complex and dense \HI~clouds (Cohen et al. 1988; Rohlfs
et al. 1984). Above 30Dor, and extending west, there is a diagonal
region in which there are only a few PNe. Since this region is
tilted at the maximum angle to our line of sight (35 degrees, see
section~\ref{section 7}), extinction may be playing a part in the
lower PNe density here. Further north, the distribution equals the
density of PNe found in the south-west region of the map.  These two
areas correspond to the opposite ends of the kinematic line of nodes
which has been shown to be twisted in the range
0$\leq$$\textit{r}$$\leq$1$^{\circ}$ and
2$^{\circ}$.8$\leq$$\textit{r}$$\leq$3$^{\circ}$.8 between
208$^{\circ}$ and 190$^{\circ}$ (Feitzinger, 1980; Rohlfs et al.
1984). We also note the relatively low numbers of newly discovered
PNe at the centre of the main bar itself where the stellar
population is at it's highest density. Intriguingly, the increasing
number density of PN towards the main bar does not continue at the
same rate onto densest stellar area. The number density of PNe
increases to the S of the main optical bar axis where the PNe are
closer in alignment to the \HI~disk (see section~\ref{section 7} for
details).

\begin{figure*}
  \includegraphics[width=1\textwidth]{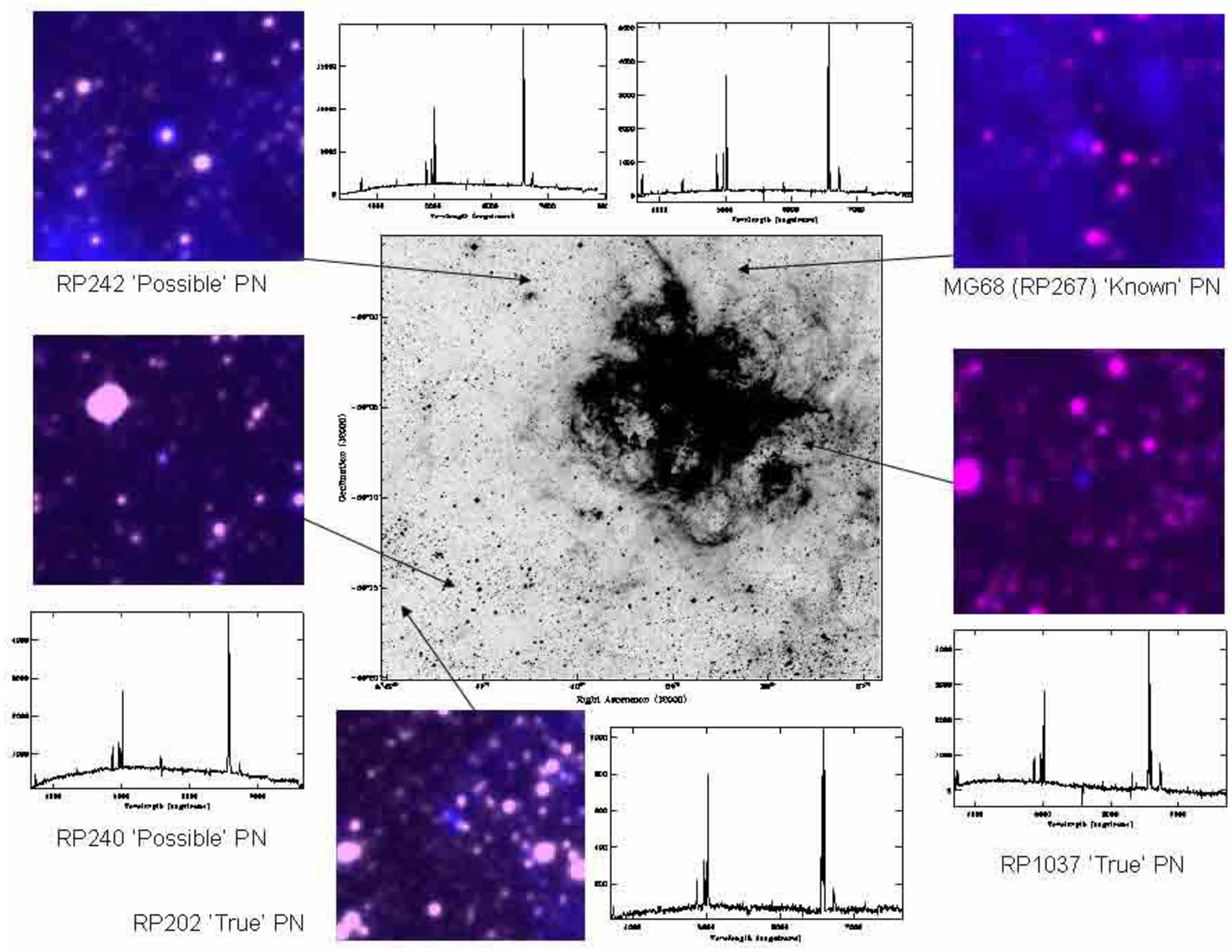}\\
  \caption{The region of 30DOR is expanded in a SR image to show previously known and new PNe with their positions, images and sky-subtracted spectra. Even in areas of dense emission, we are able to distinguish PNe by their psf and halo, coloured blue in the H$\alpha$ image.}
  \label{Figure 3}
 \end{figure*}

Several PNe were found in the region of 30Dor which is thick in
H$\alpha$ and forbidden-line emission. Sky subtraction was essential
for spectral confirmation within a 1 deg radius of 30Dor. Object
identification was achieved by reducing overall brightness while
keeping the H$\alpha$ and SR intensity levels perfectly matched.
Where candidates lie in areas between \HII~clouds and streams of
emission, we can list them as `true' otherwise they are listed as
`possible' identifications depending on the strength of the [OIII]
lines relative to H$\beta$~and the presence of \HeII~4686 (see
figure~\ref{Figure 3}).

\label{section 5}

\section{Velocities}
\subsection{Previous Surveys}

There have been seven previous kinematic studies of planetary
nebulae in the LMC. In order of publication they are Feast (1968):
25 PNe; Webster (1969): 14 PNe; Smith \& Weedman (1972): 27 PNe;
Meatheringham et al. (1988): 94 PNe; Boroson \& Liebert (1989): 68
PNe LMC \& SMC; Vassiliadis et al. (1992): 16 PNe; and Morgan \&
Parker (1998): 97 PNe. An internal dynamics study of PNe in the LMC
by Dopita et al. (1988) has largely shared the same observations and
data as the Meatheringham et al. (1988) publication. The combination
of all these studies provides a reasonable overlap of observed
objects which in total yields radial velocities for 99 PNe in the
central 25 deg$^{2}$ main bar of the LMC. Feast and Webster used
low-dispersion spectrographs and photographic plates. Feast used
three different dispersions, 49\AA mm$^{-1}$, 86\AA mm$^{-1}$ and
170\AA mm$^{-1}$, at H$\gamma$ with different cameras. Webster used
a grating giving a dispersion of 140\AA~mm$^{-1}$ in the second
order with supplementary spectrograms giving a dispersion of
50\AA~mm$^{-1}$. The probable errors ascribed to the velocity of a
single spectrograph were $\pm$5 km s$^{-1}$ for blue and $\pm$14 km
s$^{-1}$ for red. Smith and Weedman used a single-channel,
photoelectric, pressure-scanned Fabry-P\'{e}rot interferometer with
a resolution of $\sim$11 km s$^{-1}$, FWHM. A systematic error of
2.3 km s$^{-1}$ was estimated from the rms difference between pairs
of two independent sets of measurements of the same data. Measuring
the nebulae profiles in two orders, the rms difference between the
two was 3.2 km s$^{-1}$ resulting in an ultimate radial velocity
error of $\pm$5 km s$^{-1}$.

Meatheringham et al. used the 1m and 2.3m telescopes at Siding
Spring Observatory with a Perkin-Elmer \'{e}chelle spectrograph with
a resolution of 11.5 km s$^{-1}$ FWHM and a photon-counting array as
the detector. The faintest objects were then observed on the 3.9m
AAT using the Royal Greenwich Observatory spectrograph and the image
photon counting system with an overall resolution of 11.75 km
s$^{-1}$ FWHM. Their claimed errors range from $\pm$0.3 km s$^{-1}$
for narrow line profiles on the AAT to 4.8 km s$^{-1}$ for large
expansion velocities. Boroson \& Liebert used a 1200 mm grating to
yield a spectral resolution of $\sim$1.2\AA~(70 km s$^{-1}$). Morgan
and Parker used a multi-object, fibre-coupled CCD on the UKST. A
1200R grating was used with a FWHM of between 1.8 and 2 pixel across
the CCD. The 1200 line-pair mm$^{-1}$ grating produced a dispersion
of 1.34\AA~pixel$^{-1}$ in each waveband. The Morgan/Parker survey
is the only one to use the cross-correlation technique for line
measurement. Repeated velocity measurements showed a mean difference
of 2 km s$^{-1}$ and an rms scatter of 8 km s$^{-1}$. Most of the
other surveys have measured only the \OIII~line for velocity
estimates of the whole PN. In some cases (eg. Meatheringham et al.
1988) neither the measured line (believed to be \OIII~since
measurements are the same as Dopita (1988)), nor the measurement
technique has been discussed in any detail. It is therefore evident
that all past surveys have used very different means of deriving
radial velocities and calculating measurement errors for LMC PNe.

\subsection{Measurement Techniques}

Our object velocities were measured from the 2dF 1200R high
resolution spectra as described above with an estimated median
measurement accuracy of $\pm$4 km s$^{-1}$. Two different methods of
velocity measurement were employed in order to make precise
comparisons with previous published results and to expose and reduce
errors arising as a result of the application of a particular
technique.

\subsubsection{Emission Line Technique}

The {\scriptsize IRAF EMSAO} technique of measuring multiple,
specified spectral lines was first employed. Wavelengths for 13
important PN emission lines within the 6200-7350\AA~range were
specified to three decimal places. The program then applied a
weighted gaussian fit to each line dependent on its intensity,
derived a weighted average across the spectrum and corrected for the
heliocentric velocity. The {\scriptsize EMSAO} result for each PN
was displayed and examined. The program was not successful at
automatically finding spectral lines where the \NII6584\AA~line was
stronger than H$\alpha$. A manual assignment of the \NII6584\AA~line
brought all the other lines into agreement (see Figure~\ref{figure
4}). Where only 1 or 2 lines were automatically found and measured,
a manual check was made for any other missing lines that were able
to be manually measured. Extra velocities on these lines were made
using a gaussian fit in the {\scriptsize IRAF} task, splot, and
added to automatic line measurements in order to substantiate the
overall object velocity. Where lines such as OI 6300\AA~were
blended, the line was dropped from the fit and the weighted velocity
was re-calculated.

Some regular aspects of internal PN velocity structure became
noticeable. PN central stars typically eject gas with velocity
dispersions between 10 and 50 km s$^{-1}$. Even if an ejected shell
is symmetric and homogeneous, the mean velocity acquired by
measuring all the emission lines from the entire nebulae will not
represent the velocity of the central star. As expected, we found
that the H$\alpha$ line averages 6-13 km s$^{-1}$ greater velocity
than the \NII6583\AA~line. This is in part be due to the lower
excitation potential of the transition above the ground level for
the ion in H$\alpha$ compared to that for the collisionally excited
\NII6583\AA~line. The H velocities can lead to an overestimation in
the expansion velocity  due to thermal broadening (eg., Chu et al.
1984). The H$\alpha$ recombination line is able to be ionized at a
lower potential farther out, at a greater radius from the central
star, where the gas is moving at a greater velocity. This causes an
ionisation-velocity correlation with the lower excitation potential
of the H extending to greater radii depending on the temperature of
the central star. In high excitation PNe, a small increase in
velocity was noticed in the HeII~6678\AA~line where the upper level
excitation potential of the transition above the ground level for
the ion is 22.97eV. This shows that the high ionization potential
species expand at a lower velocity than the low ionization potential
species. Where excitation is high, however, inner shell high
ionisation species such as HeII~6678 will be ionised to greater
radii where there is an increase in gas velocity.

The \OI~6300 line also has an average 1-5 km s$^{-1}$ greater
velocity than the H$\alpha$ line. This line forms due to the rapid
change between H and O near the H$^{+}$ -H$^{\circ}$ ionization
front which marks the outer edge of the H$^{+}$ zone. We also found
that \OI~6363\AA~has an average 1-4 kms$^{-1}$ greater velocity than
\OI~6300\AA~where it is cleanly separated from the unshifted
\OI~6300\AA~sky line and not blended with the \SIII6312\AA~line.
Generally adjustments were small, strong H$\alpha$ lines were
automatically omitted. {\scriptsize EMSAO} recognized asymmetric
profiles which typically arise from internal nebulae inhomogeneities
and rejected such lines from the weighted fit. On no occasion were
strong high velocity lines of H$\alpha$, \OI~6300 and
\OI~6363\AA~included in our measurements. Where we had more than one
measured velocity for a PN from different observations, the velocity
which ignored the H$\alpha$ line and had the lowest errors was used.

\begin{figure}
  \includegraphics[width=0.47\textwidth]{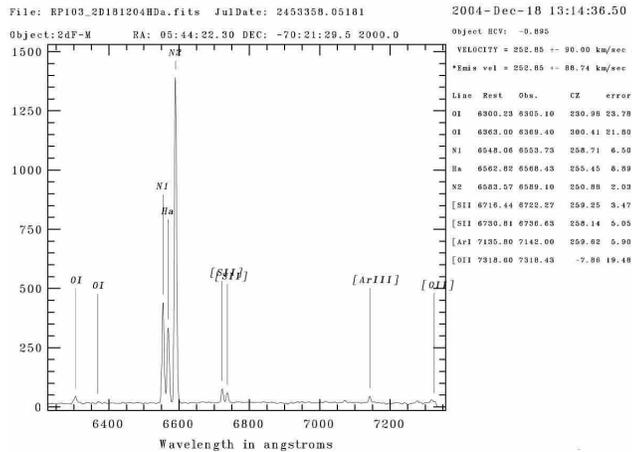}\\
  \caption{Velocities measured using the {\scriptsize IRAF EMSAO} package. A correction in identification applied to one line alone forces the whole spectrum to fit correctly.}\label{figure 4}
\end{figure}

\subsubsection{Cross-Correlation Technique}

The second method of velocity determination involved the
cross-correlation technique using {\scriptsize XCSAO} in
{\scriptsize IRAF} (Kurtz \& Mink, 1998). This method requires a
list of PN template spectra with low internal velocities and
accurately determined published radial velocities against which all
the other PN spectra may be compared for measurement. Template
emission-line velocities were based on at least four lines, and at
least two of the four must be fitted by {\scriptsize EMSAO} (Kurtz
\& Mink, 1998). Twenty templates were chosen for the
cross-correlation. Each of which was a previously known PN with
strong emission lines and variety of spectral line ratios so as to
be sensitive to the spectral variability evident in our new PN
spectra. Each template spectra also had {\scriptsize EMSAO} measured
velocities which were within $\pm$5km s$^{-1}$ of previous results
and internal line errors less than 20km s$^{-1}$. The template list
was ordered so that ratios slowly shift from high H$\alpha$/\NII6584
to high \NII6584/H$\alpha$ with varying strengths of the
\SII~doublet intermixed. This allows each input spectra to have at
least one well-matched template.

Non-matching templates produced extreme negative or high and
erroneous ($>$800km s$^{-1}$) results as a result of H$\alpha$
matching to \NII6583 and visa versa. These velocities were
eliminated while closely matching results from the remaining
spectral templates were averaged. The averaging process has the
effect of reducing the influence of any anomalies in the template
spectra velocities. The local standard of rest (LSR) at the
topographical location of the AAT and date of observation was
applied to both emission line and cross-correlation results with
reference to the position of the PN within the LMC.

\begin{figure}
  \includegraphics[width=0.47\textwidth]{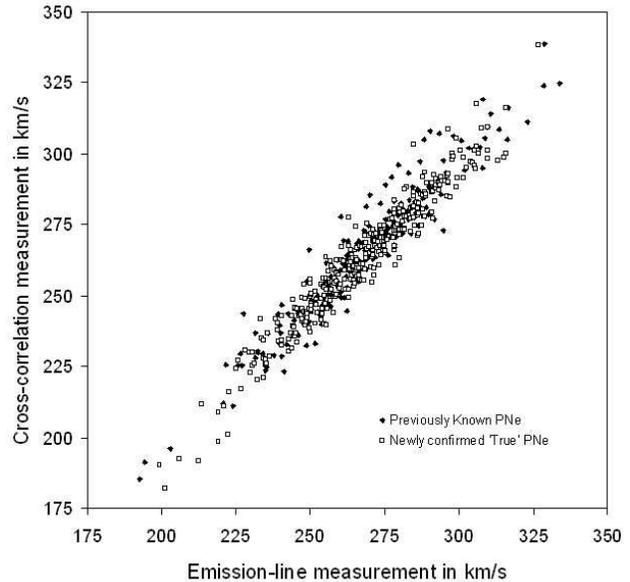}\\
  \caption{A comparison of previously known and newly discovered `true' PN velocities measured using {\scriptsize IRAF EMSAO} and {\scriptsize IRAF XCSAO}. The ratio of velocity difference is 1.02 between the two systems with a median difference of 3.94km s$^{-1}$ and an rms scatter of 6.67 km s$^{-1}$.}
  \label{figure 5}
\end{figure}

\subsubsection{Technique Comparison}

The comparison of our emission line and cross-correlation radial
velocities shows excellent agreement (see Figure~\ref{figure 5}).
The internal mean difference between our two measuring techniques is
3.9km s$^{-1}$ with an rms scatter of 6.6 km s$^{-1}$. Using
{\scriptsize EMSAO}, objects with high signal-to-noise ratios and
narrow line profiles have very low error results and include the
maximum number of lines in the weighted average. For objects with
large expansion velocities, emission line measurement errors are
greater due to the wider line profiles and greater internal velocity
structure which results in fewer lines fitted for measurement.
{\scriptsize XCSAO} fits a velocity matching the template across the
whole spectrum and therefore has many advantages over {\scriptsize
EMSAO}. Where target spectra signal-to-noise is considerably weaker
than that of the templates and fewer lines are to be found, the
cross-correlation peak may drop below 0.8 resulting in higher error
values. An average of the fitted velocity values has been derived
but where the averaged cross-correlation function (ccf) peak drops
below 0.75, it becomes preferable to use the single best fitting
template with the highest peak and lowest error. If no template with
an improved correlation and reduced error can be found, the
{\scriptsize EMSAO} velocity is used in preference.

\begin{table}
\caption{Comparison of our best measured radial velocities with
published results applying the heliocentric correction.}
{\scriptsize
 \begin{tabular}{|p{0.4cm}|p{0.3cm}|p{0.3cm}|p{0.4cm}|p{0.5cm}|p{0.4cm}|p{0.4cm}|p{0.6cm}|p{0.4cm}|p{0.3cm}|}
    \hline\hline
       \multicolumn{4}{|c|}{ \textbf{Catalog Reference}}   &   \multicolumn{1}{|l|}{\vline}              &        &     \multicolumn{3}{|c|}{\textbf{Velocity$_{helio}$}} &         \\
      \hline
      RP   &   SMP &   WS  &   J  &   RP$_{vel}$       &   B\&L &   Fea68   &   MDFW  &   S\&W &   WEB-69  \\
      \hline
10  &   88  &   27  &   ...    &   255.1  &  ...  &   308  &   226  &  ...  &   311  \\
133 &   89  &   38  &  ...  &   274.3  &  ...  &   276  &   276  &   277  &   270  \\
213 &   92  &   39  &  ...  &   279.4  &  ...  &   274  &   271  &   271  &  ...  \\
270 &   91  &  ...  &  ...  &   330.9  &  ...  &  ...  &   310  &  ...  &  ...  \\
317 &   83  &   35  &  ...  &   292.3  &  ...  &   287  &   291  &  ...  &   296  \\
400 &   60  &  ...  &  ...  &   222.7  &  ...  &  ...  &   222  &  ...  &  ...  \\
401 &   62  &   25  &  ...  &   240.7  &  ...  &  ...  &   239  &   238  &   250  \\
402 &   65  &  ...  &  ...  &   206.1  &  ...  &  ...  &   211  &  ...  &  ...  \\
404 &   71  &  ...  &  ...  &   220.8  &  ...  &  ...  &   216  &  ...  &  ...  \\
406 &   73  &  ...  &  ...  &   265.7  &  ...  &  ...  &   241  &  ...  &  ...  \\
407 &   80  &   24  &  ...  &   225.6  &  ...  &   185  &  ...  &  ...  &   187  \\
642 &   56  &  ...  &  ...  &   288.3  &  ...  &  ...  &   291  &  ...  &  ...  \\
643 &   57  &  ...  &  ...  &   310.6  &  ...  &  ...  &   313  &  ...  &  ...  \\
644 &   58  &   23  &  ...  &   294.9  &  ...  &  ...  &   279  &   276  &  ...  \\
646 &   77  &  ...  &  ...  &   236.8  &  ...  &  ...  &   343  &  ...  &  ...  \\
647 &   78  &   33  &  ...  &   262.4  &  ...  &   260  &   256  &   258  &   249  \\
648 &   82  &  ...  &  ...  &   259.8  &  ...  &  ...  &   255  &  ...  &  ...  \\
890 &   63  &   26  &  ...  &   268.5  &  ...  &  ...  &   264  &   264  &   280  \\
891 &   75  &   31  &  ...  &   304.9  &  ...  &  ...  &   301  &   297  &  ...  \\
892 &   76  &   32  &  ...  &   287.7  &  ...  &  ...  &   278  &   280  &   295  \\
1047    &   66  &  ...  &  ...  &   294.7  &  ...  &  ...  &   304  &  ...  &  ...  \\
1048    &   67  &  ...  &  ...  &   278.3  &  ...  &  ...  &   289  &  ...  &  ...  \\
1049    &   69  &  ...  &  ...  &   316.7  &  ...  &  ...  &   305  &  ...  &  ...  \\
1114    &   32  &   10  &  ...  &   254.9  &  ...  &   247  &  ...  &   257  &  ...  \\
1115    &   41  &  ...  &  ...  &   264.0  &  ...  &  ...  &   259  &  ...  &  ...  \\
1116    &   49  &  ...  &  ...  &   241.4  &  ...  &  ...  &   247  &  ...  &  ...  \\
1212    &   38  &   15  &  ...  &   255.0  &  ...  &   243  &   240  &   238  &  ...  \\
1214    &   47  &   18  &   25  &   280.6  &   274  &   275  &   272  &   269  &   280  \\
1215    &   48  &   19  &   27  &   255.2  &   251  &  ...  &   256  &  ...  &   258  \\
1216    &   51  &  ...  &  ...  &   273.5    &  ...  &  ...  &   271  &  ...  &  ...  \\
1217    &  ...  &  ...  &   5   &   272.5  &   286  &  ...  &  ...  &  ...  &  ...  \\
1218    &  ...  &  ...  &   04  &   256.7  &   256  &  ...  &  ...  &  ...  &  ...  \\
1220    &  ...  &  ...  &   12  &   243.6  &   234  &  ...  &  ...  &  ...  &  ...  \\
1221    &  ...  &  ...  &   14  &   272.0  &   251  &  ...  &  ...  &  ...  &  ...  \\
1222    &  ...  &  ...  &   15  &   232.3  &   235  &  ...  &  ...  &  ...  &  ...  \\
1223    &  ...  &  ...  &   16  &   255.2  &   253  &  ...  &  ...  &  ...  &  ...  \\
1224    &  ...  &  ...  &   17  &   249.7  &   252  &  ...  &  ...  &  ...  &  ...  \\
1227    &  ...  &  ...  &   20  &   270.1  &   271  &  ...  &  ...  &  ...  &  ...  \\
1229    &  ...  &  ...  &   22  &   260.2  &   249  &  ...  &  ...  &  ...  &  ...  \\
1230    &  ...  &  ...  &   23  &   306.3  &   247  &  ...  &  ...  &  ...  &  ...  \\
1231    &  ...  &  ...  &   24  &   260.0  &   261  &  ...  &  ...  &  ...  &  ...  \\
1233    &  ...  &  ...  &   31  &   252.5  &   260  &  ...  &  ...  &  ...  &  ...  \\
1234    &  ...  &  ...  &   32  &   271.9  &   274  &  ...  &  ...  &  ...  &  ...  \\
1235    &  ...  &  ...  &   33  &   244.2  &   244  &  ...  &  ...  &  ...  &  ...  \\
1313    &   26  &  ...  &  ...  &   253.0  &  ...  &  ...  &   256  &  ...  &  ...  \\
1395    &   28  &  ...  &  ...  &   261.4  &  ...  &  ...  &   249  &  ...  &  ...  \\
1396    &   29  &   9   &  ...  &   243.7  &  ...  &  ...  &   243  &  ...  &  ...  \\
1397    &   31  &  ...  &  ...  &   261.5  &  ...  &  ...  &   263  &  ...  &  ...  \\
1398    &   33  &   11  &  ...  &   281.0  &  ...  &  ...  &   269  &   267  &  ...  \\
1399    &   34  &  ...  &  ...  &   262.3  &  ...  &  ...  &   267  &  ...  &  ...  \\
1400    &   36  &   13  &  ...  &   261.0  &  ...  &  ...  &   263  &  ...  &  ...  \\
1401    &   37  &   14  &  ...  &   274.9  &  ...  &  ...  &   270  &   273  &  ...  \\
1403    &   42  &  ...  &  ...  &   291.4  &  ...  &  ...  &   288  &  ...  &  ...  \\
1404    &   46  &  ...  &  ...  &   271.7  &  ...  &  ...  &   273  &  ...  &  ...  \\
1405    &   52  &   21  &   34  &   276.1  &   277  &   280  &   272  &   270  &  ...  \\
1406    &   54  &  ...  &   35  &   277.6  &   276  &  ...  &   280  &  ...  &  ...  \\
1408    &  ...  &  ...  &   10  &   225.4  &   220  &  ...  &  ...  &  ...  &  ...  \\
1552    &   30  &  ...  &  ...  &   282.7  &  ...  &  ...  &   280  &  ...  &  ...  \\
1554    &   45  &   17  &  ...  &   285.8  &  ...  &  ...  &   290  &  ...  &  ...  \\
1555    &   50  &   20  &  ...  &   308.0  &  ...  &   337  &   299  &   296  &   328  \\
1556    &   53  &   22  &  ...  &   283.5  &  ...  &   277  &   277  &   278  &   300  \\
1602    &   13  &   4   &  ...  &   245.9  &  ...  &  ...  &   227  &  ...  &  ...  \\
1603    &   14  &   2   &  ...  &   266.2  &  ...  &   271  &   252  &  ...  &  ...  \\
1604    &   15  &   5   &  ...  &   191.3  &  ...  &  ...  &   203  &   202  &  ...  \\
1605    &   19  &   6   &  ...  &   239.6  &  ...  &   239  &   235  &   236  &  ...  \\
1677    &   16  &  ...  &  ...  &   253.9  &  ...  &  ...  &   253  &  ...  &  ...  \\
1679    &   18  &  ...  &  ...  &   243.1  &  ...  &  ...  &   244  &  ...  &  ...  \\
1680    &   20  &  ...  &  ...  &   287.4  &  ...  &  ...  &   288  &  ...  &  ...  \\
1682    &   24  &  ...  &  ...  &   271.4  &  ...  &  ...  &   270  &  ...  &  ...  \\
1683    &   25  &  ...  &  ...  &   208.0  &  ...  &  ...  &   188  &  ...  &  ...  \\
1797    &   21  &   7   &  ...  &   250.3  &  ...  &   250  &   259  &   262  &   283  \\
1798    &   23  &   8   &  ...  &   296.9  &  ...  &   302  &   283  &   281  &   301  \\
1894    &   27  &  ...  &  ...  &   273.0  &  ...  &  ...  &   273  &  ...  &  ...  \\
\hline
\end{tabular}}
   \label{Table 3}
 \end{table}
 \begin{table}
Abbreviations used: B\&L: Boroson \& Liebert (1989), Fea68: Feast
 (1968), J: Jacoby (1980), MDFW: Meatheringham et al. (1988) these velocities were compiled and
changed to the heliocentric system by JWC: Jacoby, Walker, Ciardullo
(1990) along with 5 of their own measurements, RP: Reid \& Parker
(this work), SMP: Sanduleak et al. (1978), S\&W: Smith, Weedman
(1972), WEB-69: Webster (1969), WS: Westerlund \& Smith (1964).
\end{table}

\begin{table}
\caption{Comparison of our best measured radial velocities with
other published radial velocities adjusted to the local standard of
rest.} {\scriptsize
  \centering
  \begin{tabular}{|p{0.3cm}|p{0.3cm}|p{0.3cm}|p{0.3cm}|p{0.2cm}|p{0.4cm}|p{0.7cm}|p{0.7cm}|p{0.7cm}|}
    \hline\hline
      &        \multicolumn{4}{|c|}{ \textbf{Catalog Reference}} &    \multicolumn{2}{|c|}{\vline} &     \textbf{Velocity$_{LSR}$}    &        \\
      \hline
RP  &   SMP &   WS  &   J &   MG  &   Mo  &   RP$_{vel}$  &   MDFW    &   MP  \\
\hline
10  &   88  &   27  &  ...  &  ...  &  ...  &   203.8   &   211.0   &  ...  \\
74  &  ...  &  ...  &  ...  &  ...  &   35  &   264.4   &  ...  &   265.5   \\
75  &  ...  &  ...  &  ...  &  ...  &   40  &   216.9   &  ...  &   248.0   \\
133 &   89  &   38  &  ...  &  ...  &  ...  &   258.9   &   261.2   &  ...  \\
152 &  ...  &  ...  &  ...  &  ...  &   37  &   247.8   &  ...  &   251.4   \\
213 &   92  &   39  &  ...  &  ...  &  ...  &   264.0   &   256.4   &  ...  \\
214 &  ...  &  ...  &  ...  &   73  &  ...  &   264.4   &  ...  &   251.6   \\
269 &   86  &  ...  &  ...  &  ...  &  ...  &   252.1   &  ...  &   260.8   \\
270 &   91  &  ...  &  ...  &  ...  &  ...  &   315.3   &   295.3   &  ...  \\
271 &  ...  &  ...  &  ...  &   75  &  ...  &   278.8   &  ...  &   282.3   \\
272 &  ...  &  ...  &  ...  &   77  &  ...  &   231.8   &  ...  &   242.0   \\
316 &  124  &  ...  &  ...  &  ...  &  ...  &   256.1   &  ...  &   245.0   \\
317 &   83  &   35  &  ...  &  ...  &  ...  &   276.3   &   276.2   &  ...  \\
399 &   59  &  ...  &  ...  &  ...  &  ...  &   221.7   &  ...  &   231.1   \\
400 &   60  &  ...  &  ...  &  ...  &  ...  &   208.3   &   207.0   &  ...  \\
401 &   62  &   25  &  ...  &  ...  &  ...  &   226.3   &   223.6   &   233.0   \\
402 &   65  &  ...  &  ...  &  ...  &  ...  &   191.3   &   195.7   &  ...  \\
404 &   71  &  ...  &  ...  &  ...  &  ...  &   205.7   &   201.2   &  ...  \\
406 &   73  &  ...  &  ...  &  ...  &  ...  &   250.6   &   225.6   &  ...  \\
407 &   80  &   24  &  ...  &  ...  &  ...  &   210.6   &  ...  &   214.6   \\
409 &  121  &  ...  &  ...  &  ...  &  ...  &   176.7   &  ...  &   207.5   \\
411 &  ...  &  ...  &  ...  &   51  &  ...  &   244.0   &  ...  &   237.0   \\
412 &  ...  &  ...  &  ...  &  ...  &   24  &   217.2   &   ...   &   207.4   \\
414 &  ...  &  ...  &  ...  &  ...  &   33  &   225.2   &  ...  &   220.8   \\
642 &   56  &  ...  &  ...  &  ...  &  ...  &   272.5   &   276.1   &  ...  \\
643 &   57  &  ...  &  ...  &  ...  &  ...  &   295.2   &   297.7   &   276.3   \\
644 &   58  &   23  &  ...  &  ...  &  ...  &   279.5   &   264.2   &  ...  \\
646 &   77  &  ...  &  ...  &  ...  &  ...  &   221.4   &   328.2   &  ...  \\
647 &   78  &   33  &  ...  &  ...  &  ...  &   247.0   &   240.7   &  ...  \\
648 &   82  &  ...  &  ...  &  ...  &  ...  &   244.4   &   239.6   &  ...  \\
649 &  117  &  ...  &  ...  &  ...  &  ...  &   258.9   &  ...  &   252.9   \\
650 &  112  &  ...  &  ...  &  ...  &  ...  &   265.5   &  ...  &   261.9   \\
651 &  116  &  ...  &  38  &  ...  &  ...  &   259.5   &   265.6   &  ...  \\
652 &  118  &  ...  &  41  &  ...  &  ...  &   206.2   &   240.0   &  ...  \\
657 &  ...  &  ...  &  ...  &   56  &  ...  &   251.6   &  ...  &   249.4   \\
658 &  ...  &  ...  &  ...  &   60  &  ...  &   247.1   &  ...  &   245.2   \\
659 &  ...  &  ...  &  ...  &   65  &  ...  &   254.1   &  ...  &   255.1   \\
661 &  ...  &  ...  &  ...  &  ...  &   27  &   248.0   &  ...  &   233.7   \\
662 &  ...  &  ...  &  ...  &  ...  &   30  &   263.5   &  ...  &   260.6   \\
890 &   63  &   26  &  ...  &  ...  &  ...  &   252.3   &   248.8   &   243.2   \\
892 &   76  &   32  &  ...  &  ...  &  ...  &   271.5   &   262.8   &   276.0   \\
893 &  ...  &  ...  &  ...  &   40  &  ...  &   210.2   &  ...  &   207.7   \\
894 &  ...  &  ...  &  ...  &   54  &  ...  &   186.9   &  ...  &   202.2   \\
895 &  ...  &  ...  &  ...  &   62  &  ...  &   257.1   &  ...  &   256.3   \\
1047    &   66  &  ...  &  ...  &  ...  &  ...  &   278.5   &   289.2   &   276.1   \\
1048    &   67  &  ...  &  ...  &  ...  &  ...  &   262.2   &   274.1   &  ...  \\
1049    &   69  &  ...  &  ...  &  ...  &  ...  &   300.7   &   289.9   &  ...  \\
1050    &  ...  &  ...  &  ...  &   42  &  ...  &   294.2   &  ...  &   297.1   \\
1051    &  ...  &  ...  &  ...  &   50  &  ...  &   272.6   &  ...  &   292.2   \\
1052    &  ...  &  ...  &  ...  &   58  &  ...  &   231.1   &  ...  &   231.8   \\
1053    &  ...  &  ...  &  ...  &   64  &  ...  &   307.5   &  ...  &   298.9   \\
1114    &   32  &   10  &  ...  &  ...  &  ...  &   240.0   &   240.2   &  ...  \\
1115    &   41  &  ...  &  ...  &  ...  &  ...  &   254.7   &   244.2   &  ...  \\
1116    &   49  &  ...  &  ...  &  ...  &  ...  &   226.4   &   232.2   &   236.4   \\
1117    &  ...  &  ...  &  ...  &   30  &  ...  &   208.3   &  ...  &   210.5   \\
1118    &  ...  &  ...  &  ...  &  ...  &   20  &   235.2   &  ...  &   247.5   \\
1212    &   38  &   15  &  ...  &  ...  &  ...  &   239.7   &   225.4   &   225.1   \\
1214    &   47  &   18  &   25  &  ...  &  ...  &   265.3   &   256.6   &  ...  \\
1215    &   48  &   19  &   27  &  ...  &  ...  &   239.1   &   240.8   &  ...  \\
1216    &   51  &  ...  &  ...  &  ...  &  ...  &   258.2   &   265.4   &  ...  \\
1217    &  ...  &  ...  &   5   &  ...  &  ...  &   257.2   &   262.9   &  ...  \\
1232    &  ...  &  ...  &   26  &  ...  &  ...  &   180.3   &   227.8   &  ...  \\
1235    &  ...  &  ...  &   33  &  ...  &  ...  &   228.2   &   231.9   &  ...  \\
1236    &  ...  &  ...  &  ...  &   28  &  ...  &   260.2   &  ...  &   275.6   \\
1237    &  ...  &  ...  &  ...  &  ...  &   19  &   259.3   &  ...  &   255.3   \\
1395    &   28  &  ...  &  ...  &  ...  &  ...  &   249.5   &   233.7   &   249.8   \\
1396    &   29  &   9   &  ...  &  ...  &  ...  &   218.0   &   228.2   &  ...  \\
1397    &   31  &  ...  &  ...  &  ...  &  ...  &   246.4   &   248.1   &  ...  \\
1398    &   33  &   11  &  ...  &  ...  &  ...  &   266.0   &   253.9   &  ...  \\
1399    &   34  &  ...  &  ...  &  ...  &  ...  &   251.7   &  ...  &   259.8   \\
1400    &   36  &   13  &  ...  &  ...  &  ...  &   245.9   &   247.6   &  ...  \\
1401    &   37  &   14  &  ...  &  ...  &  ...  &   260.3   &   255.2   &  ...  \\
1402    &   39  &  ...  &  ...  &  ...  &  ...  &   251.0   &  ...  &   214.6   \\
1403    &   42  &  ...  &  ...  &  ...  &  ...  &   278.3   &   273.4   &  ...  \\
1404    &   46  &  ...  &  ...  &  ...  &  ...  &   256.6   &   258.0   &  ...  \\
1405    &   52  &   21  &   34  &  ...  &  ...  &   259.6   &   256.9   &   259.2   \\

\hline
\multicolumn{8}{r}{continued next column $\rightarrow$}\\
 \end{tabular}}
 \label{Table 4}
 \end{table}
\begin{table}{\scriptsize
\begin{tabular}{|p{0.3cm}|p{0.3cm}|p{0.3cm}|p{0.3cm}|p{0.2cm}|p{0.4cm}|p{0.7cm}|p{0.7cm}|p{0.7cm}|}
&{\it (cont'd)}\\
\hline\hline
      &        \multicolumn{4}{c|}{ \textbf{Catalog Reference}} &    \multicolumn{2}{|c|}{\vline} &    \textbf{Velocity$_{LSR}$}     &      \\
      \hline
RP  &   SMP &   WS  &   J &   MG  &   Mo  &   RP$_{vel}$     &   MDFW    &   MP  \\
 \hline        
1406    &   54  &  ...  &   35  &  ...  &  ...  &   262.5   &   265.4   &  ...  \\
1407    &  110  &  ...  &  ...  &  ...  &  ...  &   219.7   &  ...  &   231.0   \\
1409    &  ...  &  ...  &  ...  &   19  &  ...  &   251.1   &  ...  &   268.3   \\
1410    &  ...  &  ...  &  ...  &   20  &  ...  &   268.2   &  ...  &   268.9   \\
1411    &  ...  &  ...  &  ...  &   23  &  ...  &   208.9   &  ...  &   200.5   \\
1412    &  ...  &  ...  &  ...  &   29  &  ...  &   210.7   &  ...  &   211.8   \\
1413    &  ...  &  ...  &  ...  &   31  &  ...  &   248.5   &  ...  &   259.0   \\
1552    &   30  &  ...  &  ...  &  ...  &  ...  &   267.5   &   264.9   &   263.5   \\
1554    &   45  &   17  &  ...  &  ...  &  ...  &   270.0   &   275.0   &  ...  \\
1555    &   50  &   20  &  ...  &  ...  &  ...  &   292.2   &   284.1   &  ...  \\
1556    &   53  &   22  &  ...  &  ...  &  ...  &   267.7   &   261.8   &  ...  \\
1557    &  ...  &  ...  &  ...  &   34  &  ...  &   325.8   &  ...  &   345.0   \\
1558    &  ...  &  ...  &  ...  &   35  &  ...  &   265.2   &  ...  &   276.1   \\
1602    &   13  &   4   &  ...  &  ...  &  ...  &   231.0   &   212.3   &  ...  \\
1603    &   14  &   2   &  ...  &  ...  &  ...  &   251.4   &   236.9   &   244.1   \\
1604    &   15  &   5   &  ...  &  ...  &  ...  &   176.4   &   188.0   &  ...  \\
1605    &   19  &   6   &  ...  &  ...  &  ...  &   224.8   &   220.2   &  ...  \\
1607    &  ...  &  ...  &  ...  &  ...  &   9   &   210.8   &  ...  &   203.0   \\
1608    &  ...  &  ...  &  ...  &  ...  &   11  &   230.0   &  ...  &   230.8   \\
1609    &  ...  &  ...  &  ...  &  ...  &   14  &   217.6   &  ...  &   218.1   \\
1677    &   16  &  ...  &  ...  &  ...  &  ...  &   238.7   &   237.5   &   297.1   \\
1678    &   17  &  ...  &  ...  &  ...  &  ...  &   258.0   &  ...  &   250.1   \\
1679    &   18  &  ...  &  ...  &  ...  &  ...  &   227.9   &   228.7   &   233.9   \\
1680    &   20  &  ...  &  ...  &  ...  &  ...  &   272.2   &   273.1   &  ...  \\
1682    &   24  &  ...  &  ...  &  ...  &  ...  &   256.3   &   254.7   &  ...  \\
1683    &   25  &  ...  &  ...  &  ...  &  ...  &   177.4   &  ...  &   175.0   \\
1685    &  107  &  ...  &  ...  &  ...  &  ...  &   251.4   &  ...  &   274.6   \\
1686    &  ...  &  ...  &  ...  &   14  &  ...  &   246.8   &  ...  &   248.6   \\
1687    &  ...  &  ...  &  ...  &  ...  &   12  &   244.9   &  ...  &   242.9   \\
1688    &  ...  &  ...  &  ...  &  ...  &   13  &   230.7   &  ...  &   237.3   \\
1689    &  ...  &  ...  &  ...  &  ...  &   16  &   216.9   &  ...  &   207.9   \\
1797    &   21  &   7   &  ...  &  ...  &  ...  &   234.3   &   243.7   &  ...  \\
1798    &   23  &   8   &  ...  &  ...  &  ...  &   281.4   &   268.0   &  ...  \\
1799    &  106  &  ...  &  ...  &  ...  &  ...  &   272.3   &  ...  &   280.9   \\
1800    &  ...  &  ...  &  ...  &   8   &  ...  &   287.8   &  ...  &   275.4   \\
1801    &  ...  &  ...  &  ...  &   10  &  ...  &   247.9   &  ...  &   254.9   \\
1802    &  ...  &  ...  &  ...  &   15  &  ...  &   260.6   &  ...  &   262.5   \\
1894    &   27  &  ...  &  ...  &  ...  &  ...  &   256.6   &   258.2   &  ...  \\
1895    &  ...  &  ...  &  ...  &   9   &  ...  &   245.6   &  ...  &   245.0   \\
1896    &  ...  &  ...  &  ...  &   11  &  ...  &   201.1   &  ...  &   206.9   \\
\hline
 \end{tabular}}
 \\[1cm]
 Abbreviations used: J: Jacoby (1980), MDFW: Meatheringham et al. (1988), MG: Morgan \& Good (1992), Mo: Morgan (1994), MP: Morgan, Parker (1998), RP: Reid \& Parker (this work), SMP:
Sanduleak et al. (1978), WS: Westerlund \& Smith (1964).
 \end{table}

\begin{table}
\caption{Mean velocity differences for LMC PNe between previously
published LSR and heliocentric velocities and the corresponding LSR
and heliocentric velocities obtained in this survey. Separate
results for both the EMSAO and XCSAO measurement methods are given
with the standard deviation.}{\scriptsize
\begin{tabular}{|c|c|c|c|c|cl}
\hline\hline
  Survey &     EMSAO & Std. Dev. &   XCSAO & Std. Dev.& No. \\
          & mean diff. &        & mean diff. &   &  PN \\
          &(km s$^{-1}$) & (km s$^{-1}$) &  (km s$^{-1}$) & (km
          s$^{-1}$) &  \\
  \hline
  RP-XC & 0.7 & 6.2 &  - &  - & 130  \\
  MDFW  & 1.3 & 7.7  &  3.9  &  12.2 & 58 \\
  MP    & -0.5  &  11.3  &  -0.6 &  11.7 & 71 \\
B\&L    &  1.9  &  7.5  & 1.1  &  10.5 & 19\\
Fea68  &   1.2  &  10.1  &  0.2  &  11.6 & 21 \\
S\&W  & 6.0  &  8.4  &   7.3  &  10.4 & 25 \\
Web-69  &  -1.4  &  6.6  &  1.6  &  13.2 & 17 \\
 \hline
 \end{tabular}}
 \label{Table 5}
   \\[1cm]
Abbreviations used: B\&L: Boroson \& Liebert (1989), Fea68: Feast
(1968), MDFW : Meatheringham et al. (1988), MP: Morgan, Parker
(1998), RP: Reid, Parker (this work), S\&W: Smith, Weedman (1972),
WEB-69: Webster (1969).
 \end{table}

\subsection{Comparison with Previous Results}

\begin{figure}
  \includegraphics[width=0.48\textwidth]{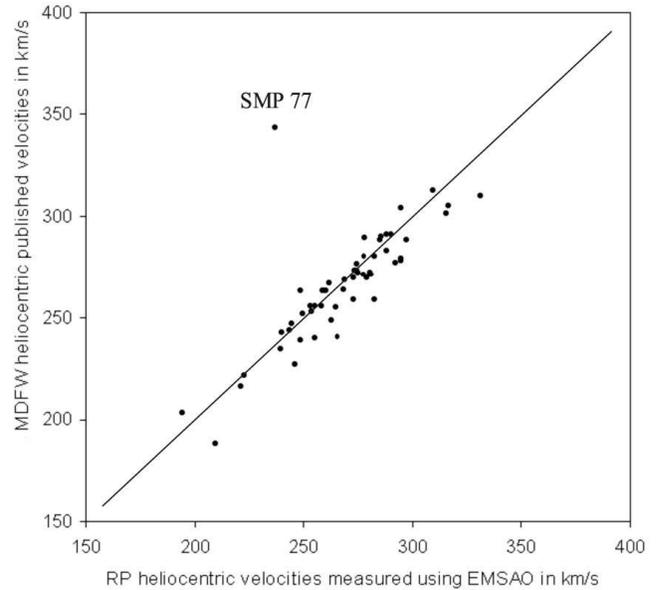}\\
  \caption{A comparison of the MDFW heliocentric velocities with our emission line measurements. The object in the upper left is SMP77 which is $\sim$100 kms$^{-1}$ greater than our measurement. We believe the MDFW velocity to be erroneous.}
  \label{figure 6}
 \end{figure}

\begin{figure}
  \includegraphics[width=0.48\textwidth]{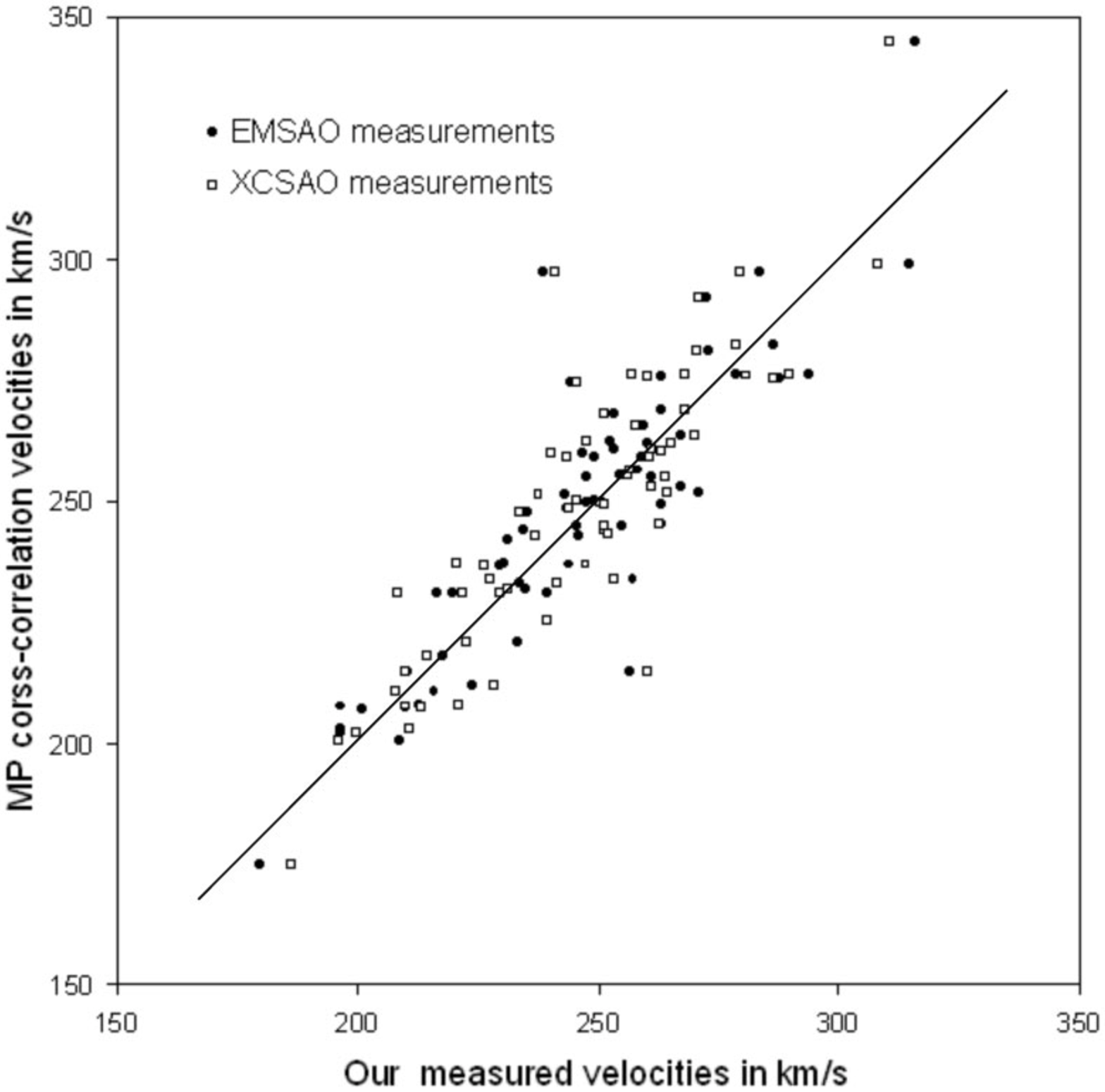}\\
  \caption{A comparison of our velocities with the MP velocities, both measured using the cross-correlation technique (open boxes). Our corresponding emission line measurements are included (filled angle circles) showing that the emission line method of measurement produces equal or very close results in most cases. The diagonal line represents the position where the RP velocity measurements equal MP velocity measurements. }
  \label{figure 7}
 \end{figure}

A comparison of our heliocentric and LSR radial velocities measured
using EMSAO with previously published results is shown graphically
in figures~\ref{figure 6} and \ref{figure 7}. Agreement is generally
very good with the exception of SMP 77, seen at the upper left in
figure~\ref{figure 6}. For this PN we have derived an emission line
heliocentric velocity of 237.01 km s$^{-1}$ and a cross-correlated
heliocentric velocity of 231.85 km s$^{-1}$. Since the MDFW result
is almost 100km s$^{-1}$ greater than our measurements, we suspect
their value may be a typographical error.

Multiple observations of PNe using different 2dF field plate
configurations has resulted in 382 and 223 duplicated
cross-correlation and emission-line measurements respectively. Mean
differences in velocities between these observations vary by less
than 3 km s$^{-1}$ with an rms scatter of 5.6 km s$^{-1}$. After
examining all multiple observations for any single object, we have
chosen to use the velocity with the smallest error values.

\subsection{Our Adopted Radial Velocity Estimates}

To derive a best radial velocity from our emission line and
cross-correlation methods, we examined the error and other
properties relating directly to each measurement system. In the
emission line technique, we looked for a large proportion of fitted
lines applied to each spectrum. In addition we sought overall error
values $\leq$23 km s$^{-1}$ where errors larger than this value
begin to result from increasingly complex internal velocity
structure. Error values up to 23 km s$^{-1}$ through this technique
are to be expected, as internal velocity ratios between different
lines vary with standard regularity as described above. In the
cross-correlation technique, we looked for high correlation peaks
and low error values $\leq$2 km s$^{-1}$. For our accepted
velocities, we used the result with the lowest error according to
each measurement system. Emission line fitted results were used
where the cross-correlation technique was not successful, or where
errors were above 23km s$^{-1}$. The cross-correlation technique was
used where too few emission lines were measured with {\scriptsize
EMSAO} or no weighted result was possible.

Our best derived velocities from both the emission line and
cross-correlation methods are directly compared with other published
velocities. Table~\ref{Table 3} compares our heliocentric
measurements with five earlier determinations. Table~\ref{Table 4}
compares our measurements, converted to the local standard of rest
(LSR), with other previous velocity determinations which have been
published with only the LSR correction. As shown in summary
Table~\ref{Table 5}, there is excellent agreement with results from
Feast (1968), hereafter (Fea68) and Meatheringham (1988):(MDFW). The
velocities of MDFW represent the largest sample for heliocentric
comparison within our survey area, with 58 PNe. The mean difference
using emission line measurements is 1.3 km s$^{-1}$ for MDFW and
only 1.2 km s$^{-1}$ for Fea68, although the standard deviation for
Fea68 is the highest at 10.1 km s$^{-1}$. Using the
cross-correlation measurements, the mean difference with Fea68 is
the smallest value we found, at only 0.2 km s$^{-1}$. The standard
deviation however increases to 11.6 km s$^{-1}$. The results from
Smith and Weedman's 1972 survey (S\&W) have the largest mean
difference compared to both of our measurement techniques. The
standard deviation within that lower range however, is in keeping
with the other results, indicating a systemic error or offset in the
S\&W data of about 5 km s$^{-1}$. The Webster (1969):(Web-69)
velocities show the highest standard deviation when comparing our
cross-correlation technique results. This survey, with only 17 PNe
within our survey area, is also the smallest with which to make a
comparison.

We have converted our best velocities to the local standard of rest
in order to make direct comparisons with two previous surveys;
Morgan and Parker (1998)(MP) and Meatheringham et al. (1988)(MDFW).
The individual PN results are given in Table~\ref{Table 4} with the
mean differences shown in Table~\ref{Table 5}. Here, the MP survey
is of particular interest as their UKST fibre spectral velocities
were derived using the cross-correlation technique. The mean
difference using emission line measurements between our survey and
MP is only -0.5 km s$^{-1}$ and -0.67 km s$^{-1}$ when compared to
our cross-correlation results. The MDFW mean difference of 1.3 km
s$^{-1}$ is also in good agreement when compared to our emission
line measurements. Unlike the MP results however, this difference
increases to 3.9 km s$^{-1}$ when compared to our cross-correlation
measurements. Nevertheless, all of these comparisons indicate
excellent agreement. The mean difference between our emission line
results and previously published results is 1.41 km s$^{-1}$ with
$\sigma$=1.2 km s$^{-1}$. Similarly for the same comparison using
cross-correlation results the mean difference is 2.25 km s$^{-1}$
with $\sigma$=1.95 km s$^{-1}$. Taking the overall adopted radial
velocities from both the emission line measurements and the
cross-correlation measurements, we find a mean difference of 1.87 km
s$^{-1}$with a standard deviation of 9.8 km s$^{-1}$. This indicates
an overall agreement better than 2 km s$^{-1}$ with previous LMC PN
radial velocity surveys.

\label{subsection 6.5}
\label{section 6}


\section{PNe Kinematics within the LMC}

\subsection{History and current practice}

The study of PNe in the LMC affords us a unique opportunity to
investigate the connection between the late stages of stellar
evolution and the dynamics of the interstellar medium (ISM) in
galaxies. In the past, the main method for studying the ISM have
been observations of \HI.
The first full LMC structure surveys by McGee (1964) and McGee \&
Milton (1966) were followed by more detailed studies by Rohlfs et
al. (1984) and Luks \& Rohlfs (1992) both using the Parkes radio
telescope. An \HI~synthesis survey using the Australia Telescope
Compact Array (ATCA) was conducted by Kim et al. (1998) and provided
a detailed structure with which to correlate surface brightness
density of \HI~gas with \HII~regions and objects such as PNe and
SNRs. Several surveys of the \HI~distribution have shown that much
of the \HI~structure and surface densities are correlated with giant
and supergiant shells identified in H$\alpha$ emission. The
\HI~distribution has an axisymmetric appearance as opposed to the
optical appearance of the LMC. \HI~also has a velocity range
V$_{hel}$ 190-387 km s$^{-1}$ (Kim et al. 1998).

In more recent times, other specific objects have been used to trace
out LMC kinematics such as PNe (Meatheringham et al. 1988) and
carbon stars (van der Marel, 2001) and core helium-burning red clump
stars (Olsen \& Salyk, 2002). Without the long held assumption that
the LMC disk is circular, van der Marel (2001), using the Two Micron
All Sky Survey (2MASS) near-IR data, has produced a new and accurate
description of LMC geometry and dynamics. The method uses pure
geometry, relying on the sinusoidal variations in the apparent
brightness of the LMC as a function of its position angle on the
sky. The estimated viewing angles are $\textit{i}$ = 34$^{\circ}$.7
$\pm$6$^{\circ}$.2 and $\Theta$ = 122$^{\circ}$.5 $\pm$8$^{\circ}$.3
(defined that the near side of the LMC is at
$\Theta$$_{near}$$\equiv$$\Theta$--90$^{\circ}$). The distribution
of carbon stars are viewed as preferable to \HI~since the use of
carbon stars (van der Marel, 2001) places the dynamic stellar centre
at the optical centre of the bar. Their study shows that the shape
of the LMC disk is actually elliptical and has a nonuniform surface
density distribution, indicative of tidal forces, interacting from
our galaxy and the SMC.

\begin{figure*}
  \includegraphics[width=1\textwidth]{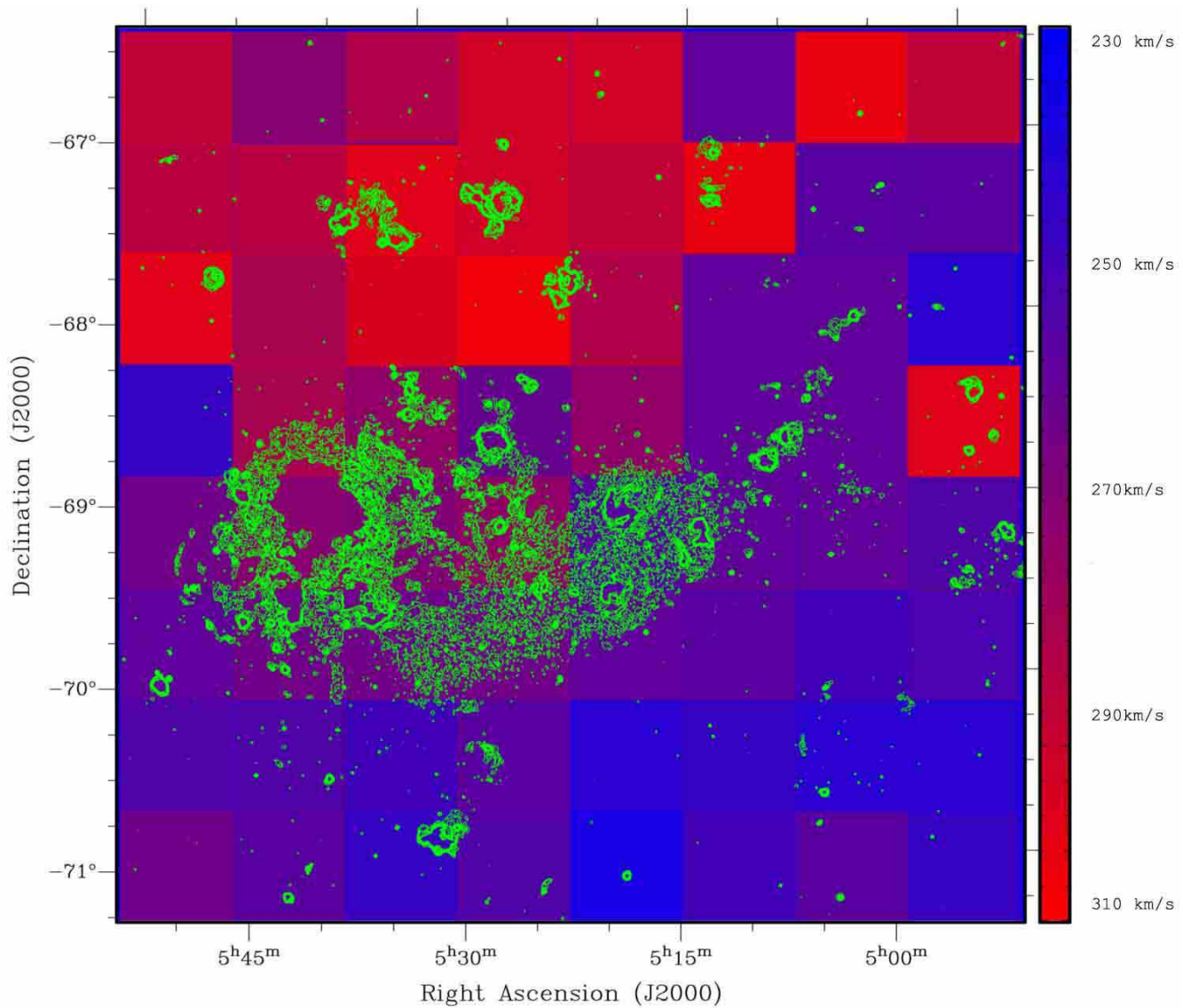}\\
  \caption{A PNe radial velocity map of the central 25 deg$^{-2}$ of the LMC coloured from blue to red with increasing heliocentric velocity. Velocities are averaged for all PNe within each 37 $\times$ 37 arcmin area where each image cell is divided into 4 equal quadrants. The map is overlayed in H$\alpha$ image contours (30-80\% intensity) in order to more easily examine velocity profiles and extinction. NE is to the top left hand corner.}
  \label{figure 8}
 \end{figure*}

\subsection{PN vs \HI~Radial Velocity Distribution}

With this knowledge, we compare our large new sample of LMC PNe
velocities with established \HI~velocities. This comparison reveals
tidal disruption in the LMC system. In order to observe the mean PN
velocity distribution across the central bar region, we have
averaged our adopted velocities within 37 $\times$ 37 arcmin sub
regions. This has the effect of countering the effects of individual
PNe which may have large peculiar velocity motions. A map of this
velocity distribution (figure~\ref{figure 8}) shows an overall shift
in the velocity of the PN population running NE (RA. 05h 45min Dec.
-67deg) to SW (RA. 5h 00min Dec. -71deg). Some perturbation of the
PN velocities at either end of the main bar (indicated by strong red
and blue colours side by side) could be the result of a mild
systemic, rotation of the PN population. In the ``kinematic circular
disk method" of tracing LMC dynamics, Meatheringham et al. (1988)
assume a circular disk movement to find the position angle $\Theta$
of the kinematic line of nodes. In a true circular galaxy, this line
of maximum velocity gradient will represent the true line of nodes.
With the advantage of our additional PN data however we can see that
the line representing the maximum velocity gradient (an imaginary
line between red and blue colour cells in Figure~\ref{figure 8})
cannot be represented by a straight line. It has a curved movement
from the NW to the centre of the main bar and then somewhat upwards
to the NE. In their study of LMC proper motion, HIPPARCOS (Kroupa et
al. 1997) indicated that the LMC may be rotating in a clockwise
direction on the sky. This may account for part of this observed
velocity gradient. In addition, our 37 $\times$ 37 arcmin averaged
velocities confirm that the PNe population in the northern part of
the disk, above the bar, is rotating away from us. The approaching
south-eastern arm of the LMC also appears to be the nearest side to
us (Klein et al 1993; van der Marel, 2001).

\begin{figure*}
  \includegraphics[width=0.65\textwidth]{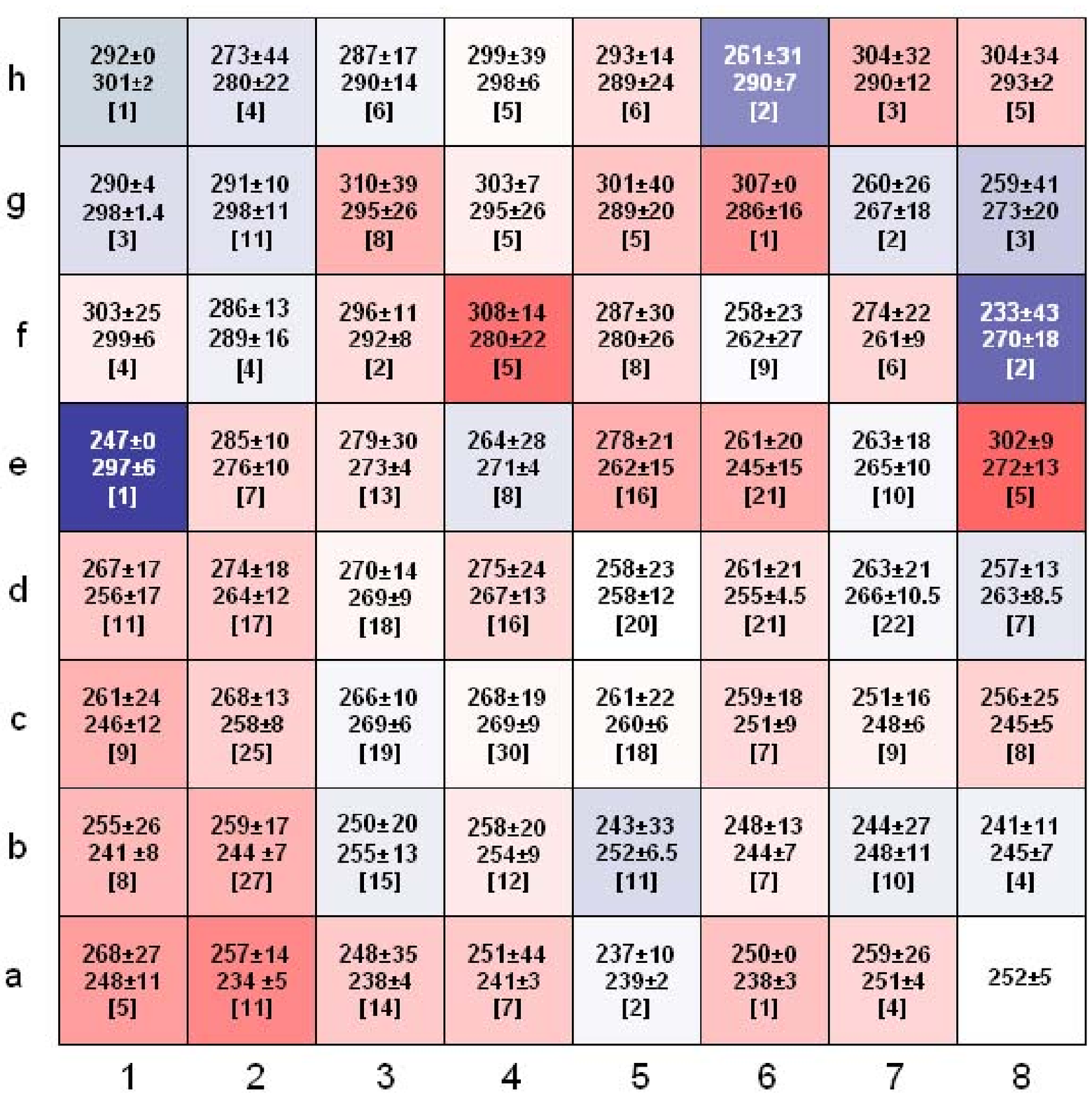}\\
  \caption{A PNe versus \HI~radial velocity map of the central 25 deg$^{2}$ of the LMC giving the mean average PNe and \HI~heliocentric velocities within each $\sim$37 $\times$37 arcmin cell. The PN velocity is given first with the standard deviation of velocities within that cell. Underneath we give the mean average \HI~velocity (Rohlfs et al. 1984) with the standard deviation from 12 pointings within the same cell. Within square brackets we give the number of PNe in each cell. Each cell is coloured according to the movement of the averaged PNe with respect to the \HI~gas. The intensity of the colour increases with the difference in velocity. PNe moving at greater velocity than the \HI~disk are coloured red and those moving at a lesser velocity are coloured blue. Positions correspond exactly to the map in figure~\ref{figure 8}. }
  \label{figure 9}
\includegraphics[width=0.7\textwidth]{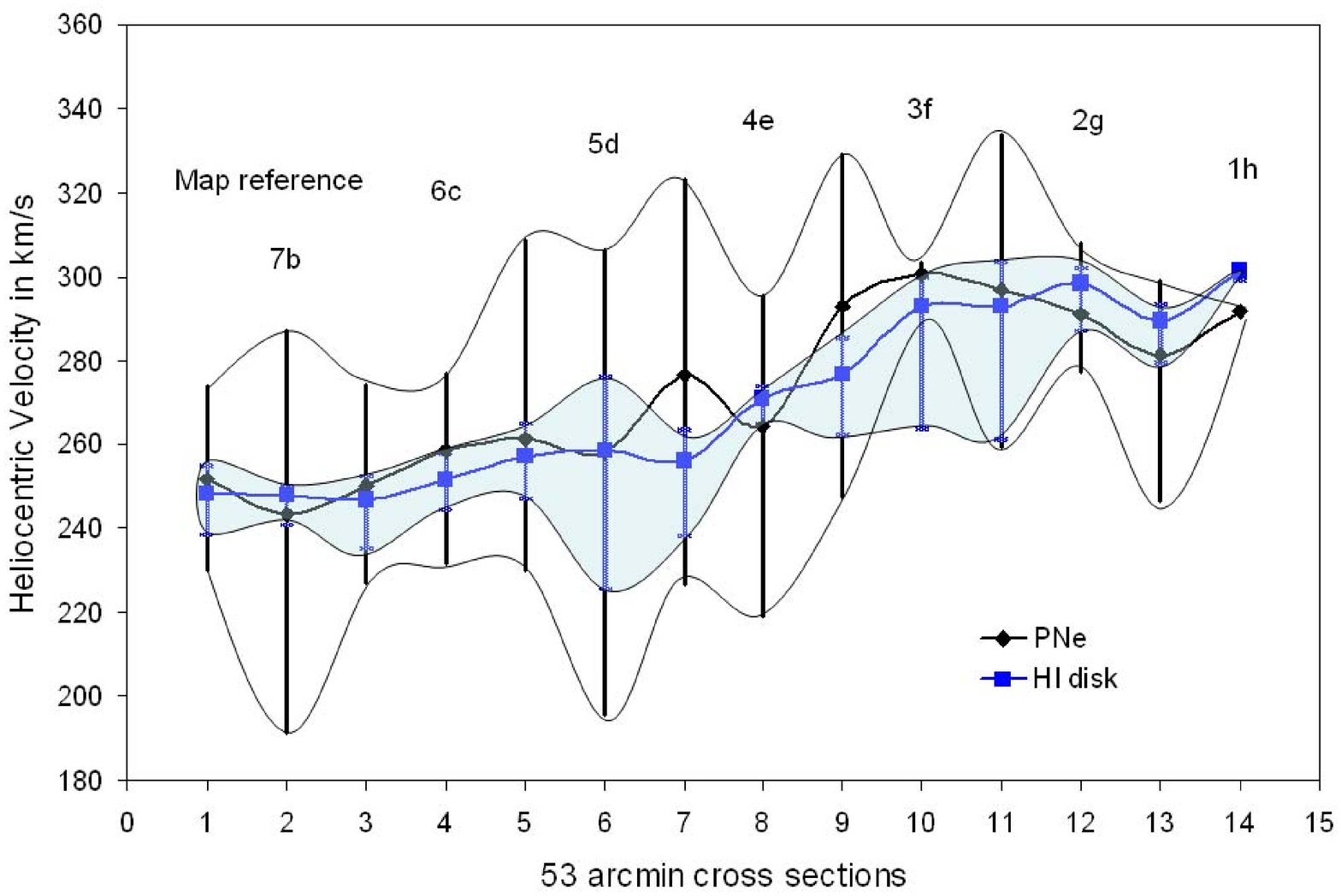}\\
\caption{A heliocentric vertical dispersion graph comparing PNe and
the \HI~disk perpendicular to the line of nodes (Figure~\ref{figure
9} map ref. 7a-1h). This clearly reveals the LMC's angle of
inclination to our line of sight. Averaged velocities for PNe and
\HI~within cross-sections, each 53 arcmin apart and 2 degrees wide,
are shown as solid points, joined by a curved trend line. The shaded
blue area shows the full range of \HI~velocities within a given
cross-section. The black error bars, likewise show the full range
above and below the average PN velocity for a given cross-section.}
 \label{figure 10}
 \end{figure*}

We have compared our heliocentric PN velocities with the \HI~
velocities of Rohlfs et al. (1984) and find that the central PN
population resides at an angle consistent (but not constant) with
the \HI~disk. The large \HI~cloud embedding 30Dor and extending
south $\sim$2 deg is divided into high and low velocity components.
Our averaged PN disk fits almost centrally between both of these
\HI~clouds. Averaged PNe in this region are slightly greater in
velocity than other PNe fields to the immediate west, beyond the
main \HI~clouds. This may be an indication that the older stellar
population was less perturbed by the LMC's close encounter with our
galaxy than was the \HI~disk.

In Figure~\ref{figure 9} we show the same area as pictured in
Figure~\ref{figure 8}, however we give the averaged PN velocity and
averaged \HI~velocity for each sub-region along with the standard
deviation for each. Regions here are coloured red to blue with
varying intensity which relates to the difference in velocity
between the PN and \HI~data for each cell. The number of PNe in each
cell is given in square brackets. Cells g6, h1, e1 and a6 are
represented by only 1 PN. Cell a8 contains no PNe so only the
\HI~velocity is given. Overall, the velocity dispersion of the PN
population is greater than that of the \HI~disk. The comparison
shows that both populations are somewhat perturbed at either end of
the main bar in a NW to SE direction, close to the line of nodes.
The faint colouring of cells in the SE of the grid indicates a near
parallel average movement of the 2 populations in this area.

Figure~\ref{figure 10} shows a cross-section along the angle of
inclination, perpendicular to the line of nodes (given by an
imaginary line, 2 degrees wide joining map references 7a-1h). The
slice is divided into sections, each 53 arcmin apart. The velocities
are based on the average heliocentric velocities shown in
Figure~\ref{figure 9}, however here we give the full range of the
dispersion for both the PN population and the \HI~disk. The maximum
dispersion is shown by a shaded colour for the \HI~disk and joined
error bars for the PN population. The slice clearly reveals the
LMC's angle of inclination to our line of sight. The \HI~disk has
the widest dispersion at cross-section 6, where it corresponds to
the widest PN velocity dispersion and the densest stellar section of
the main bar. The size of 53 arcmin for the cross sections was
arrived at by dividing each sub-cell in half across a 45 deg angle
in line with the direction of the slice.

The PN population in the NE of the map (above 30Dor) with map
reference h-g; 1-2 are moving at considerably greater velocity than
the kinematic centre of the population but generally at a lower
velocity than the \HI~gas in that region. Based on $\Delta$V, the
mean PNe population in that area (h-g; 1-2) is angled at
4$^{\circ}$.5 to the \HI~disk. In the NW corner of the map (ref.
f-h; 6-8), there is considerable mixing and interweaving of the
averaged PNe population and \HI~disk, with high velocity dispersions
indicated by the high standard deviation values.

The typical radial velocity dispersion for \HI~within each specified
sub-cell is only $\sim$18 km s$^{-1}$, but that increases to an
average of 40 km s$^{-1}$ above the main bar where the steepest
velocity gradient lies. The measurement and comparison of PNe and
\HI~indicates that the innermost 2.6 deg bar region has non-circular
motions. We may therefore conclude that the LMC disk is warped and
its line of nodes is twisted to the NW. This may produce the twists
in the SW section of the line of nodes observed by Olsen \& Salyk
(2002). This conclusion needs to be checked and examined in a more
sophisticated manner.

The observed velocities of these PNe however is not the true
velocity of these PNe as part of the LMC field. The large angular
extent of the LMC requires correction for the change in velocity
from one side to the other. In order to derive the rotation
characteristics for the central PN population, the effect of solar
motion must be removed. The distance modulus of the LMC has been
estimated to be 18.5 $\pm$0.1 (Feast 1984; Visvanathan 1985). The
distance from the sun to the LMC is $\delta$ = 50.1 $\pm$2.3 kpc.
The radial velocity of the LMC (LSR) is 250$\pm$5 km s$^{-1}$. With
the IAU adopted V$_{\odot}$=220 $\pm$7 km s$^{-1}$ standard orbital
motion of the sun about the Galactic centre, (Einasto, Haud,
J$\hat{o}$eveer 1979) we can use the value of the transverse
velocity of the LMC as seen at the sun (V$_{obs}$) to derive the radial velocity (V$_{r}$) of the LMC as seen from the sun.\\

V$_{r}$=V$_{obs}-V_{\odot}$(sin $\delta$ sin $\delta_{\odot}$ + cos
$\delta$ cos $\delta_{\odot}$ cos $\Delta\alpha$) \\

The origin of the coordinate system is established in the centre of
each of our 64 sub-cell areas as shown in Figure~\ref{figure 8}. In
this case $\Delta\alpha$ is the difference between the RA at the
centre of the LMC and the position of the PN within a particular
sub-cell. The effective solar apex is taken to be $\alpha_{\odot}$=
18h 03m 50s, $\delta_{\odot}$= 30deg 00m 16s.

The projected shape and kinematics of the LMC is determined by the
angles through which it is viewed (the inclination angle
$\textit{i}$ and the position angle $\Theta$ of the line of nodes).
For this purpose, we adopt $\textit{i}$ = 35.8 deg $\pm$2.4 deg and
$\theta$ = 145 deg $\pm$4 deg (Olsen \& Salyk, 2002). To obtain this
measurement, Olsen \& Salyk used a least squares fit of a plane to
their core helium-burning red clump stars over 50 fields across the
whole galaxy. They also removed 15 fields in the SW edge which were
subject to the tidal influences of our galaxy. Their value for
$\textit{i}$ is in good agreement with van der Marel \& Cioni
(2001), who found a value of $\textit{i}$ = 34.9$\pm$6.2 deg.

\begin{figure}
  \includegraphics[width=0.49\textwidth]{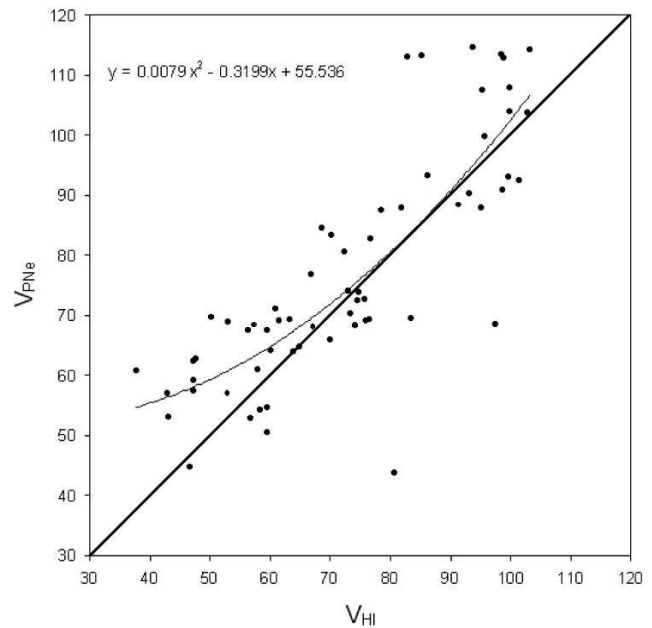}\\
  \caption{The averaged GSR velocity for PNe in km s$^{-1}$ compared with the averaged \HI~velocity within each cell. The solid line represents the line of slope unity where the PN population would equal the \HI~velocity. The faint line represents the least squares best fit between PNe and \HI~velocities with the associated equation. }
  \label{figure 11}
 \end{figure}

From this geometry, using the averaged PNe V$_{h}$ at the centre of
the optical line of nodes (264 km s$^{-1}$), we find a
galactocentric velocity of 70.7 km s$^{-1}$ for the central LMC PNe
population. This may be compared to a velocity of 71 km s$^{-1}$ at
the kinematic point of symmetry using \HI~data (Rohlfs et al. 1984).
This point of coincidence is located at RA 05 19 12 DEC -69 26 17.
The dispersions for the PNe and \HI~disk are so wide at this point
that we cannot claim that this is a point of symmetry for both
populations. If we accept the slow and solid rotation of the main
bar, then the lack of a single point of symmetry for the PN
population will not effect the viewing angles $\textit{i}$ and
$\theta$. To compare the PN and \HI~dynamics, we compare the
averaged PNe and averaged \HI~velocities within each cell, adjusted
to the galactic standard of rest. This is shown in
Figure~\ref{figure 11} and should be compared to Figure 4 in Freeman
et al. (1983) and Figure 6 in Meatheringham et al. (1988).

Although the deviations for the PNe and \HI~gas disk within each
cell have not been included here, the averaged radial velocities
show a general trend. At both low and high \HI~velocity levels, the
PN population have considerably higher velocities than the \HI~gas
plane. The faint line in Figure~\ref{figure 11} represents the
polynomial least squares fit (with 3$\sigma$ rejection) between the
PN population and the \HI~disk. The solid line represents the line
of equality, where the PN velocity would equal the \HI~velocity. The
diagram shows a closer correlation between the PN and \HI~dynamics
than that found by Meatheringham et al. (1988). The reason for this
lies in our use of larger numbers of PNe averaged in small cells and
the accuracy of our velocity measurements. This averaging reduces
any peculiar velocity effects from individual PN as previously
explained.


The PN population is angled further away from our line of sight in
the north east of the LMC main bar. In this area they also become
embedded behind the \HI~disk. This however raises the question of
whether the younger PN population reside closer to the \HI~disk. In
order to answer this question we are working on the abundance ratios
for our new PN sample and will present results relating to the
kinematics of Type I PNe in a further paper in this series.

\label{subsection 7.2}
\subsection{Velocity Gradients}

\begin{figure}
  \includegraphics[width=0.49\textwidth]{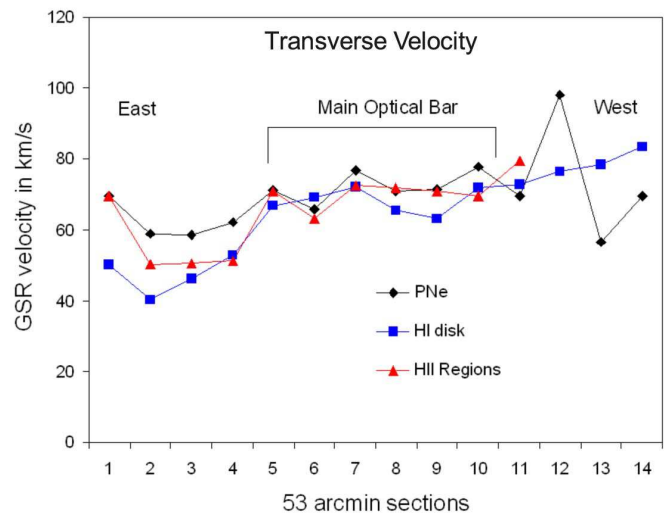}\\
  \caption{A comparison of the transverse velocity for PNe, \HI~gas and \HII~regions, averaged in 53 arcmin sections along the line of nodes at $\Theta$ = 145 deg. Each section strip, 5 deg in width, running from Figure~\ref{figure 9} coordinates 8f-1b, represents an average of the GSR velocity for each population. The area of closest stability between PNe, \HII~regions (including previously known and compact new \HII~regions) and the \HI~gas disk is seen to occur along the main optical bar. Sections 3 - 5 containing the 30 Dor region show that the \HI~disk is more perturbed than the PN population.}
  \label{figure 12}
 \end{figure}

\subsubsection{Transverse Velocity}

Without assuming circular motion of the LMC, after correction for
the transverse motion of the LMC (subsection~\ref{subsection 7.2}),
the kinematic line of nodes for the PN population has a twist of
$\sim$20 deg to the NW of the main bar. The transverse velocity
gradient across the central bar has therefore been investigated by
adopting $\Theta$ = 145 deg (Olsen \& Salyk, 2002) for \HI~and
creating perpendicular sections, each 5 deg wide, across the survey
region. The sections are spaced 53 arcmin apart and represent the
averaged velocities for PNe, \HII~regions and \HI~gas respectively.
The \HII~region velocities have been measured using the same methods
described in section~\ref{section 6} as part of the entire H$\alpha$
emission-line survey of the central 25 deg$^{2}$ of the LMC. Many of
these \HII~regions are our new compact discoveries. As seen in
figure~\ref{figure 12}, gradual increase in velocity E to W is
detectable along the line of nodes on the main optical bar. Further
to the east, the lower velocity of the \HI~disk is due to tidal
interaction with our galaxy and the resulting material drawn toward
the Magellanic stream. The asymmetric position of the bar
(Feitzinger 1983) may also be observed through the \HI~gas streaming
motions to either side. West of the main bar, the PN population
depart quickly from any relative \HI~association. This is strong
evidence for a solid-body rotation of the main bar to 2.6 deg either
side of the centre of symmetry.

We may conclude that the main bar represents the most stable region
of star formation and has been the least perturbed part of the LMC
following its near collision with the SMC which may have occurred
$\sim$2 $\times$ 10$^{8}$ years ago (Murai and Fujimoto, 1980). To
either side of the main bar we witness the strong tidal effects on
the exponential disk (Freeman, 1970). Similar effects have been
observed in the SMC (Mathewson and Ford, 1984; Dopita et al. 1985;
Mathewson et al. 1986).

\subsubsection{Angle of Inclination}

\begin{figure}
  \includegraphics[width=0.472\textwidth]{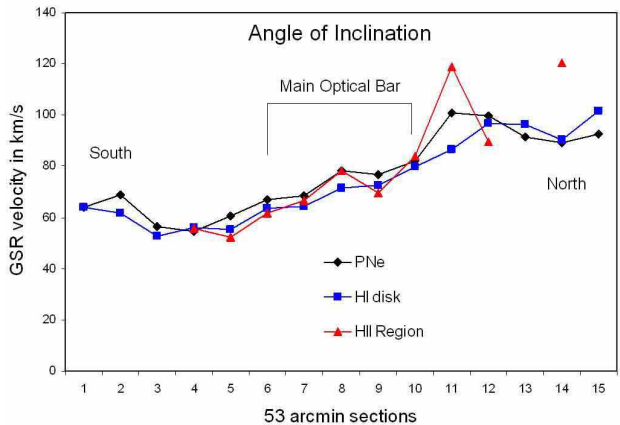}\\
  \caption{A comparison of GSR velocities for PNe, \HI~and \HII~regions averaged in 53 arcmin sections along the angle of inclination, perpendicular to the line of nodes. Each section strip, 5 deg in width running from Figure~\ref{figure 9} coordinates 8a-1h, represents an average of the GSR velocity for each population. Cross-sections correspond to the same positions given in Figure~\ref{figure 10}.}
  \label{figure 13}
 \end{figure}

We have investigated and compared the averaged velocity gradients
for PNe, \HI~gas and \HII~regions in 58 arcmin sections moving from
SW to NE across our 25 deg$^{2}$ survey area. Each section is 4.3
deg in width and crosses the main bar region from positions 6 to 9
shown in figure~\ref{figure 13}. The gradual increase in the
\HI~velocity gradient is direct evidence of the 35 deg angle of
inclination (Olsen \& Salyk, 2002). Like the gradient examined along
the line of nodes, we also find here that the PN and \HI~disk
correspond very closely across the face of the bar. To either side
however we can see evidence of slight warping which is probably due
to the combined effects of the tidal interaction between the solid
body rotation of the main bar itself with the exponential disk and
past interaction with the SMC.

\label{section 7}

\subsection{Main bar rotation curves for \HI~and planetary
nebulae}

Two rotation curves have been obtained from the averaged PN and \HI~
velocities. Each curve is taken from a strip $\sim$1.5 deg wide,
divided into 1.5 deg sections. The strips have been labeled
`position 1' and `position 2' indicting their different positions
across the bar region. The strip at position 1 follows the high
velocity edge of the kinematic line of nodes for the PN population.
The strip at position 2 follows the centroid of the PN distribution
on the main bar and passes 0.5 deg north of the \HI~centroid (Rohlfs
et al. 1984). It then moves to the NW, following $\Theta_{max}$
rather than the $\Theta$ as shown in Figure~\ref{figure 12}.
Previous results from \HI~and discrete traces have shown
$\Theta_{max}$ - $\Theta$ = 20 - 60 deg (van der Marel 2001b).
Averaged GSR velocities used for the rotation curves have been
adjusted for the transverse velocity of the LMC
(section~\ref{subsection 7.2}) and are shown in Figure~\ref{figure
14}. The solid rotation of the main bar is indicated by the
perturbation of both the \HI~and PN curves between cells 4 and 7 in
position 2. It has previously been suggested that the bar is offset
from the centre of the outer disk by $\sim$0.5 deg (Westerlund
1997). The suggestion has also been made that the bar may reside in
a separate plane to the outer disk (Zhao \& Evans 2000), but we see
little evidence for this.

\begin{figure}
  \includegraphics[width=0.472\textwidth]{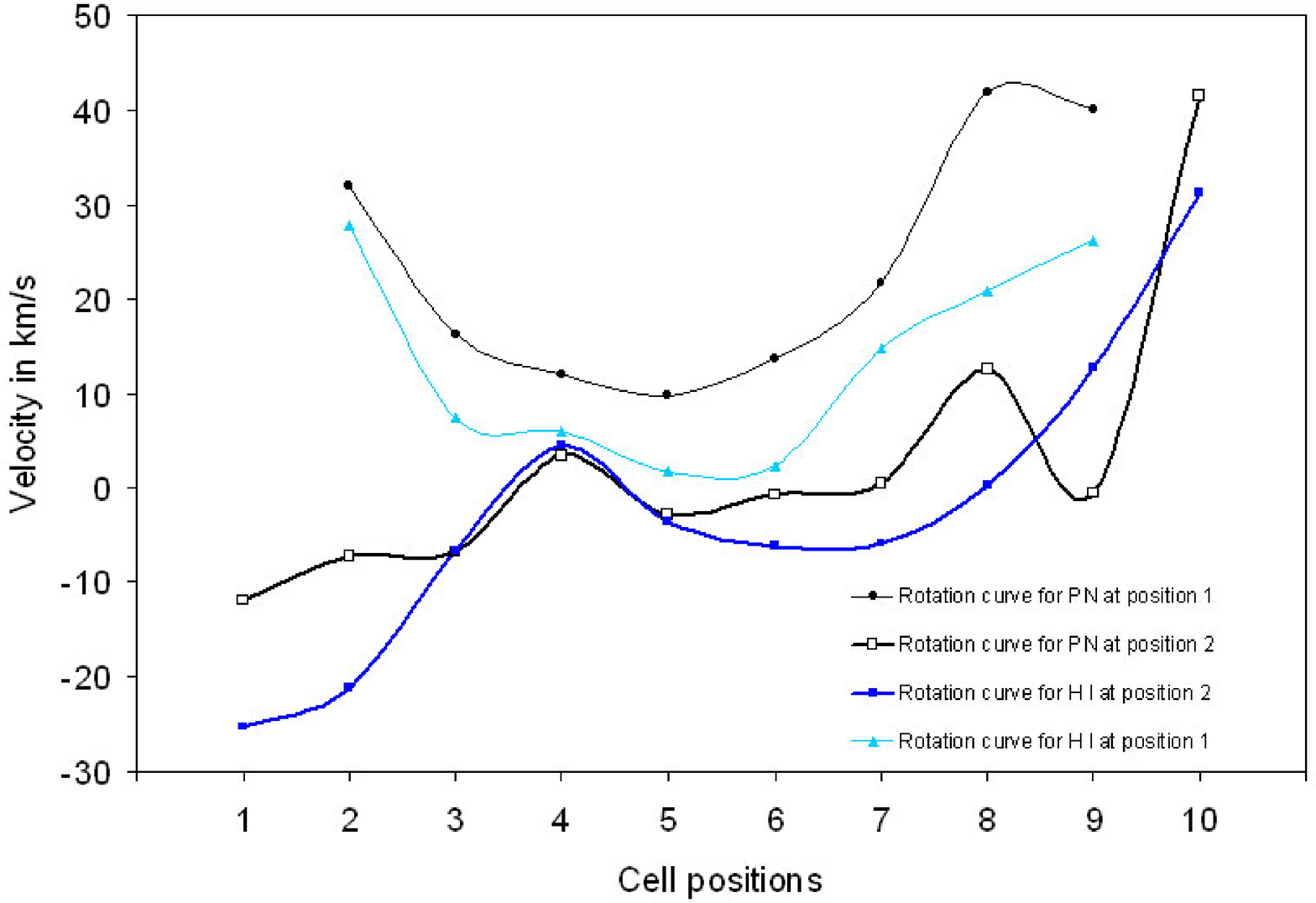}\\
  \caption{Two LMC main bar rotation curves, each comparing the velocities of PNe and \HI~(Rohlfs et al., 1984) across $\sim$1.5 deg wide strips.  The slower, solid rotation of the main bar is evident between cells 4 and 7 on the position 2 rotation curve. }
  \label{figure 14}
  \includegraphics[width=0.472\textwidth]{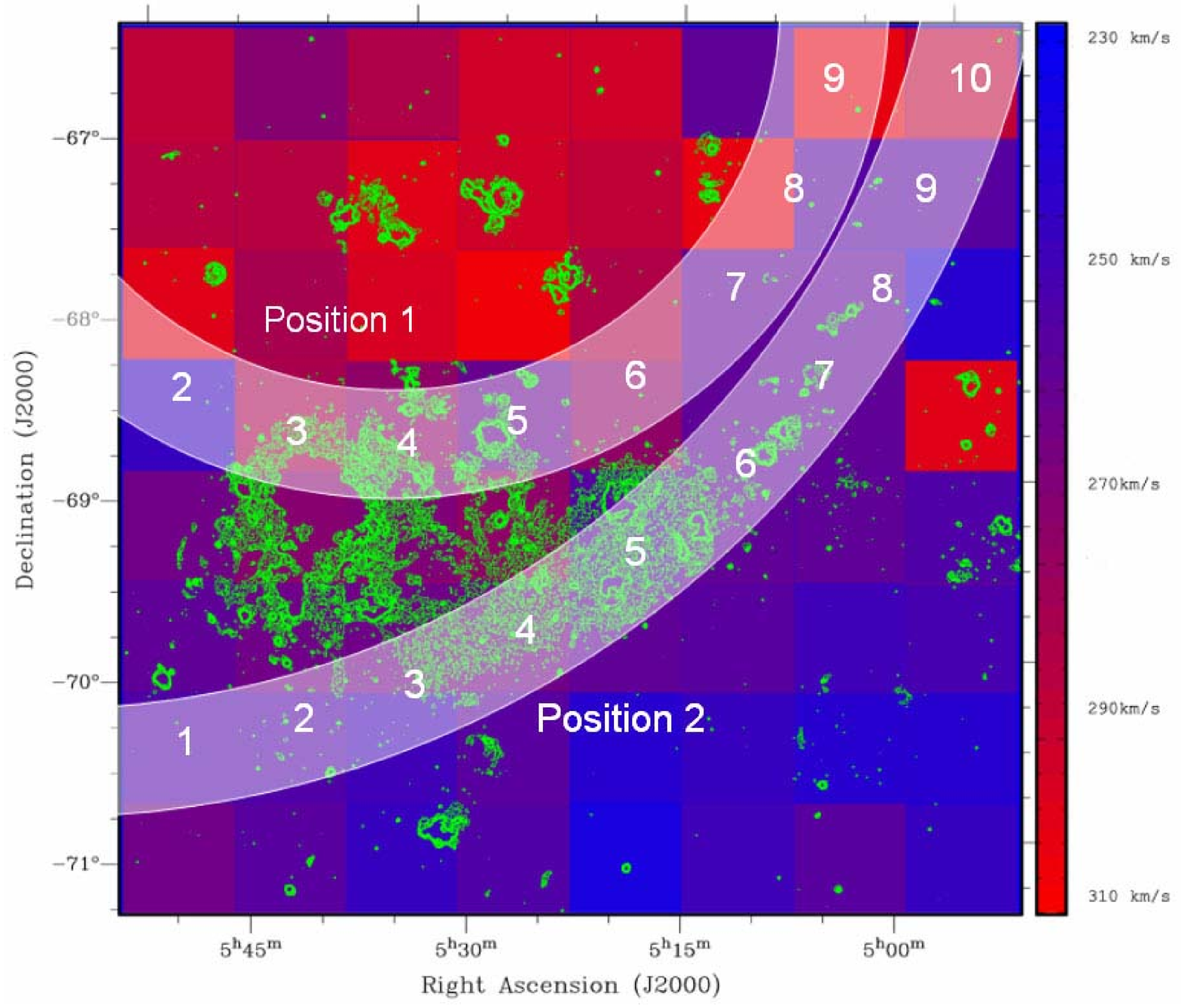}\\
  \caption{A heliocentric velocity grid of the LMC indicating the positions for the two rotation curves shown in Figure~\ref{figure 12}. }
  \label{figure 15}
 \end{figure}

Figure~\ref{figure 15} shows the positions of both strips while the
background heliocentric velocity map provides an indication of the
increased velocity across the region. The high velocity strip at
position 1 turns to the NE after crossing the northern half of 30
Dor. Other than this change of direction there appears to be little
else about the 30 Dor region that has any particular effect on
either the \HI~disk or the PN velocities. It has been stated that 30
Dor is not within the LMC disk and may be the centre of it's own
\HI~velocity field (Luks \& Rohlfs, 1992). This separate `L
component' velocity field of Luks \& Rohlfs has been examined using
21-cm absorption and found to be on the far side of the LMC disk
(Dickey et al. 1994). This is consistent with the regional stability
of our PNe velocities across the observed face of 30 Dor.

It is interesting to compare the curves derived from both positions.
In the high velocity strip (position 1), the average PN velocity is
consistently greater than the average \HI~velocity. To the south,
the PNe follow the \HI~disk but with increased perturbation once
they have departed from main bar. This immediately indicates that
the probability of finding young PN is higher on the main bar. The
curvature of $\Theta_{max}$ to the NW may also be an indication of
the ellipticity of the LMC which has been found to be as great as
$\epsilon$'' = 0.312 $\pm$0.007 in the outer disk (van der Marel
2001). This distortion of the disk is probably the result of
interaction with our galaxy presently and the SMC mainly in the past
(van der Marel, 2001). It is possible that the bar itself can be
formed by interactions (Gerin et al. 1990; Barnes 1998). The
disruption of the bar, seen in both the \HI~disk and PNe velocities
in the position 2 curve may be caused by a lopsided mass
distribution (Rudnick \& Rix, 1998; Swaters et al., 1999). It is
therefore possible that the shift to positive velocities as the main
bar is approached in cells 4 \& 5 may be due to the underlying mass
of 30 Dor. Overall, it is evident that the LMC is a complex rotating
system which lacks the standard symmetry usually observed in spiral
galaxies.


\section[]{The New LMC PN Catalogue}

\subsection{Previously Known PNe}

All previously known PNe within the 25 deg$^{2}$survey region were
independently recovered during our search and have been included in
Table~\ref{table 6}. They are listed in order of ascending RA and
Dec (columns 3 \& 4) along with the RP catalogue number (column 1)
and other common identifications (column 2). New, updated positions
were determined from our calibrated stacked H$\alpha$ map and double
checked against the online SuperCOSMOS maps (Hambly et al. 2001) so
that all LMC PNe, new and previously known are on the same uniform
astrometric grid.

The diameters (column 5) including the extended halos have been
measured on the H$\alpha$ maps using {\scriptsize GAIA} in the
{\scriptsize STARLINK} package. The point spread of compact,
star-like objects grows as a function of their apparent magnitude
and exposure time on the tech-pan film. Without saturation, however,
the UKST is able to resolve any extended object with a diameter
greater than 4 arcsec. The smallest object size we can measure
covers 5 pixels. With 0.67 arcsec/pixel data, this gives a diameter
of 3.3 arcsec. Most of the diameters are measured on a complete
pixel to pixel basis so that they represent multiples of 0.67,
however, where we detect faint illumination in an outer pixel, we
approximate the extra distance within that pixel. Conversion of
diameters to distances in parsec have been provided in column 6.
These are based on an LMC distance of 50kpc $\pm$3kpc (Mould et al.
2000) where each arcsec corresponds to $\sim$0.25pc. For more
details on measured diameters please see RP1.

The tabulated velocity (column 7) is our ascribed measurement as
described in section~\ref{section 6}. The velocity errors differ in
their meaning according to the method of measurement. For velocity
measurements based on emission lines (em), the error (column 8)
reflects internal variations within a given PN's own individual line
velocities. For LMC PNe, errors $\leq$23km s$^{-1}$ are within the
expected range; especially where 6 or more lines are included. For
velocity measurements based on cross-correlation (ccf), the error
estimator (column 9) is derived analytically with the additional
assumption of sinusoidal noise, whereby the half-width of the
sinusoid is equal to the half-width of the correlation peak. The
derived error estimator is
\begin{displaymath}
error = \frac{3}{8}~  \frac{\textit{w}}{1 + \textit{r}},
\end{displaymath}
where $\textit{error}$ is the error in a single velocity measurement
by xcsao, $\textit{w}$ is the FWHM of the correlation peak, and
$\textit{r}$ is the ratio of the height of the true peak to the
average peak (Tonry \& Davis 1979). Errors $\leq$2.5 km s$^{-1}$ are
preferable because larger errors indicate the inclusion of
non-matching templates.

Comments (column 10) have been abbreviated so that c=circular
(round), e=elliptical (oval), f=faint, b=bright, s=small, p=point
source objects which are small in the short red image,
ireg=irregular, ds= double star, bp= bipolar. Bipolarity cannot be
clearly confirmed from the 0.67 arcsec resolution of the H$\alpha$
map however, large, elliptical PNe with strong \NII/H$\alpha$ ratios
are strong contenders. A comment of `fading' refers to the outer
appearance of the halo in H$\alpha$. Most PNe have a well defined
outer boundary however several gradually fade in a manner analogous
to most emission line stars. A comment of `H$\alpha$ only' means
that the object is well defined in H$\alpha$ and only faintly
visible in the short red (if at all). The comment c$\sim$e shows
that while the object is circular, there appear to be some interior
components that tend to be elliptical.

Only four previously known PNe in the survey region were not
allocated a fibre on the crowded 2dF plate and therefore have no
velocity by our measurement. More detailed comments on individual
objects will be presented in our next paper in this series where
line diagnostics will be presented. Four previously known PNe have
been commented as `possible PN' or one as a `likely PN' and 2 as
`Not PN'. These comments have been made after close examination of
the object images and spectra. It is possible that some may fit into
a class of very low excitation (VLE) PNe which could be very young
PNe with cool stellar temperatures $\textsl{T$_{z}$}$(H)
$\sim$31,500K (Meatheringham \& Dopita 1991). J19 (RP1226), however
has both the appearance and spectrum of a variable star. No emission
lines were visible and no velocity was measurable from this object's
spectrum, therefore we conclude it is not a PN.



\subsection{Newly Discovered PNe}

All the newly discovered PNe have come from the central 25deg$^{2}$
central bar region of the LMC. They were identified as small
emission sources using the digitized stack of 12 deep, high
resolution (0.67 arcsec) H$\alpha$ images, as described in
section~\ref{section 1}. Further details on the procedure are given
in RP1.

\begin{figure}
  \includegraphics[width=0.47\textwidth]{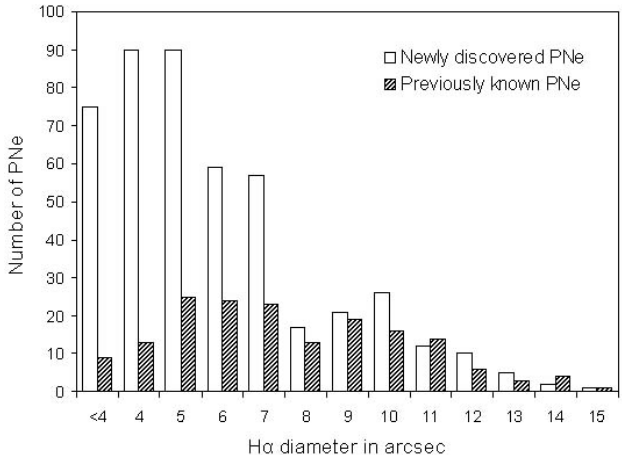}\\
  \caption{H$\alpha$ diameter versus number for both previously known and newly discovered PNe with the whole 25 deg$^{2}$ central bar region of the LMC. The small angular size of much of the new sample is clear. Many PNe of comparable size to the previously known, however, were also found although most of them are somewhat fainter.}
  \label{figure 16}
 \end{figure}

Our large program of spectroscopic follow-up has revealed 460 new
PNe in our survey area. Our newly identified PNe cover the same size
ranges as the previously known sample but they extend the ratio of
small and faint nebulae as might be expected due to our deep
H$\alpha$ map (see Figure~\ref{figure 16}. There are several PNe,
however, discovered by Jacoby (1980) and Morgan (1994) which are
comparable in size and shape but at a higher luminosity. We found a
large number of PNe which were visible only in H$\alpha$. The red
image has no definite counterpart but often tiny stars are present
in the field and may or may not be associated with the PN that was
found in that position.

In Table~\ref{table 7} we list all the newly discovered PNe in order
of their RA. The first column gives the RP catalogue reference
number. In all, this catalogue contains over 2,000 emission objects
which were initially numbered as they were discovered east to west
across the survey area. Column 2 gives the IAU representative name
to assist in object follow-up observations. Columns 3 and 4 give the
right ascension and declination of each new PN in J2000
co-ordinates. In column 5 we give the H$\alpha$ measured diameter of
the object in arcsec. The apparent diameters of bright objects
$\leq$4 arcsec will suffer from a psf growth as a function of
exposure time and object luminosity when saturated (RP1). The
diameters are nonetheless useful as an initial indication of object
size and/or brightness. The heliocentric radial velocity with the
lowest errors is given in column 6 with the error in column 7. The
meaning of the error result is dependant upon the method of velocity
measurement. The emission-line measurement (em) error indicates the
weighted average difference in the objects' own internal velocities.
The cross-correlation (ccf) error given in column 8, is the averaged
error across all matching templates used to produce the velocity.
The object probability (P) is summarized in column 9 where we have
abbreviated true PNe as `T', likely PNe as `L' and possible PNe as
`P'. Column 10 is an abbreviated list of optical comments. The
nomenclature is similar to that already described for the previously
known PNe with the addition that 44 PNe are VLE objects. They are
generally below 19th mag in H$\alpha$ and have intensities
\OIII~5007 $\leq$ H$\beta$. We have included them as possible PNe
due to the lack of significant continuum, the presence of other
nebulae emission lines such as \OII~3727 and their optical
appearance. More work will be required to positively identify these
objects as either a peculiar population of faint LMC PNe or as
extremely compact, circular \HII~regions. In column 10, the phrase
`H$\alpha$ only' is a reference to the appearance of the nebulae on
the merged SR and H$\alpha$ map. These objects are not directly
visible in the SR so only the assigned colour for the H$\alpha$
image is visible. In all, 218 new PN are visible in H$\alpha$ only.
One object, RP1088, has been detected entirely by it's
\NII~emission. This PN may be a member of the rare H-deficient group
of PN like Abell 30 and Abell 78 (Jacoby \& Ford 1983). Spectroscopy
has confirmed that there is almost no H$\alpha$ emission from this
PN. A paper on this object is in preparation.

We compare the diameters of the previously known central LMC PNe
with the new PNe presented here. The previously known PNe, including
their outer extended halos, have a mean diameter of 7.6 arcsec, a
median diameter of 7.3 arcsec and a standard deviation of 2.7
arcsec. The new PNe have a mean diameter of 6.0 arcsec, a median of
5.3 arcsec and a standard deviation of 5.3 arcsec. Clearly, the new
PNe are generally both smaller and fainter than the majority of
those previously known. See figure~\ref{figure 16} for a graphical
representation of the diameter comparison.

\begin{table*}
\caption{Newly determined velocities and accurate positions for all
previously known PNe within the central 25deg$^{-2}$ region of the
LMC.}
\begin{tabular}{|l|l|c|c|c|c|c|c|c|l|}
\hline \hline
Cat. Ref. &   Other Cat. Ref. &   RA    &   DEC   &    Dia  &  Dia & V  &   V$_{err}$ &   V$_{err}$  &   Comments    \\
          &                & J2000   &   J2000 &  arcsec &  pc  & helio          &  em  &   ccf   &    \\
\hline
RP1607  &   Mo9     &   04 52 11.39 &   -70 12 46.4    &   4.0 &   0.97    &   225.7   &       &   2.5 &   e, p    \\
RP1606  &   MG4     &   04 52 45.24 &   -70 17 49.1    &   7.4 &   1.79    &       &       &       &   e, s    \\
RP1800  &   MG8     &   04 55 39.79 &   -68 34 20.3    &   9.4 &   2.28    &   303.3   &   20.3    &       &   e   \\
RP1895  &   MG9     &   04 56 21.09 &   -67 24 22.5    &   5.2 &   1.26    &   261.4   &   25.4    &       &   e, s, p \\
RP1801  &   MG10    &   04 56 37.79 &   -67 40 54.6    &   5.8 &   1.41    &   263.4   &   22.0    &       &   e, s, p \\
RP1608  &   Mo11    &   04 58 03.23 &   -70 25 08.8    &   6.2 &   1.50    &   244.9   &       &   1.2 &   e, p, in \HII   \\
\end{tabular}
 \label{table 6}
 \end{table*}

\begin{table*}
\begin{tabular}{|l|l|c|c|c|c|c|c|c|l|}
&{\it (cont'd)}\\
\hline \hline
Cat. Ref. &   Other Cat. Ref. &   RA    &   DEC   &   Dia  &  Dia & V  &   V$_{err}$ &   V$_{err}$   &   Comments    \\
          &                & J2000   &   J2000 &  arcsec & pc & helio          &  em  &  ccf    &    \\
\hline
RP1403  &   SMP42   &   05 15 46.81 &   -68 42 23.7    &   6.5 &   1.58    &   293.4   &   23.0    &       &   e, s    \\
RP1223  &   J16   &   05 15 54.68 &   -69 30 35.7    &   6.7 &   1.62    &   255.2   &       &   1.8 &   c, s, H$\alpha$ only    \\
RP1413  &   MG31    &   05 16 29.29 &   -68 18 11.3    &   4.0 &   0.97    &   263.6   &   16.2    &       &   c, s, H$\alpha$ only    \\
RP1224  &   J17   &   05 17 00.72 &   -69 19 29.1    &   4.0 &   0.97    &   249.7   &   23.7    &       &   c, b, p \\
RP1213  &   SMP43   &   05 17 02.46 &   -69 07 16.4    &   6.0 &   1.45    &   260.5   &   17.3    &       &   e, p    \\
\end{tabular}
 \end{table*}

\begin{table*}
\begin{tabular}{|l|l|c|c|c|c|c|c|c|l|}
&{\it (cont'd)}\\
\hline \hline
Cat. Ref. &   Other Cat. Ref. &   RA    &   DEC   &    Dia  &  Dia &  V  &   V$_{err}$ &   V$_{err}$   &   Comments    \\
          &                & J2000   &   J2000 &  arcsec  & pc & helio          &  em  &  ccf    &    \\
\hline
RP404   &   N207 SMP71  &   05 30 33.30 &   -70 44 37.6    &   11.0    &   2.67    &   220.8   &   23.3    &       &   c, lop-sided    \\
RP405   &   SMP72   &   05 30 45.83 &   -70 50 15.8    &   7.3 &   1.77    &   240.8   &       &   1.1 &   H$\alpha$ to north  \\
RP406   &   N208 SMP73  &   05 31 21.97 &   -70 40 44.9    &   15.0    &   3.64    &   265.7   &   18.2    &       &   distorted   \\
RP662   &   Mo30    &   05 31 35.30 &   -69 23 46.4    &   10.7    &   2.59    &   278.9   &   18.8    &       &   e, poss bp  \\
RP657   &   MG56    &   05 31 45.00 &   -69 43 07.2    &   6.0 &   1.45    &   267.0   &       &   1.7 &   e, H$\alpha$ only, in \HII  \\
\hline
\end{tabular}
  \\[1cm]
Explanation of abbreviations used: BE: Bohannan \& Epps (1974), J:
Jacoby (1980), ~LI: Lindsay (1963), ~LM1: Lindsay \& Mullan (1963),
~LM2: Lindsay \& Mullan (1963), ~MG: Morgan \& Good (1992), ~Mo:
Morgan (1994), ~N: Henize (1956), ~RP: Reid \& Parker (this work,
2006), ~SMP: Sanduleak et al. (1978), ~Sa: Sanduleak (1984), ~WS:
Westerlund \& Smith (1964)
\\
\end{table*}

\begin{table*}
\caption{Velocities and accurate positions for all newly discovered
PNe within the central 25deg$^{-2}$ region of the LMC.}
\begin{tabular}{|l|l|c|c|c|c|c|c|c|l|}
\hline \hline
Cat. Ref &   IAU Name &   RA    &   DEC   &  H$\alpha$ Dia  &   V   &   V$_{err}$ &  V$_{err}$  &   P  &   Comments    \\
        &            &      J2000  &   J2000  &   arcsec  & helio    & em   &   ccf  &     &     \\
\hline
RP1695  &   RPJ 045233-700043   &   04 52 33.97 &   -70 00 43.7    &   11.5    &   267.6  &       &   1.4  &   T   &   e, b    \\
RP1758  &   RPJ 045345-695758   &   04 53 45.08 &   -69 57 58.7   &   10.0    &   273.3  &      &   1.6  &   T   &   c, b    \\
RP1636  &   RPJ 045353-704455   &   04 53 53.72 &   -70 44 55.5    &   5.0 &   249.9  &   18.4   &     &   T   &   e, f, H$\alpha$ only    \\
RP1714  &   RPJ 045403-693320   &   04 54 03.56 &   -69 33 20.8    &   5.0 &   278.0  &       &   1  &   T   &   c, b, s, p  \\
RP1716  &   RPJ 045424-692942   &   04 54 24.49 &   -69 29 42.5    &   27.0    &   255.0  &   21.0   &     &   P   &   e, b central zones  \\
RP1783  &   RPJ 045433-692035   &   04 54 33.78 &   -69 20 35.7    &   7.4 &   240.1  &       &  2.6  &   P   &   c, s, b, known els, VLE \\
\end{tabular}
 \label{table 7}
 \end{table*}

\begin{table*}

\begin{tabular}{|l|l|c|c|c|c|c|c|c|l|}
&{\it (cont'd)}\\
\hline \hline
Cat. Ref &   IAU Name &   RA    &   DEC   &  H$\alpha$ Dia  &   V   &   V$_{err}$ &   V$_{err}$    &   P  &   Comments    \\
        &            &      J2000  &   J2000  &   arcsec  & helio    & em   &  ccf   &     &     \\
\hline
RP1624  &   RPJ 050531-710206   &   05 05 31.10 &   -71 02 06.1    &   7.0 &   234.3  &       &   1  &   T   &   c, f, diffuse, H$\alpha$ only   \\
RP1649  &   RPJ 050551-703443   &   05 05 51.60 &   -70 34 43.1    &   2.2 &   242.1  &       &   1.1  &   P   &   e, s, p \\
RP1719  &   RPJ 050600-692830   &   05 06 00.37 &   -69 28 30.1    &   3.0 &   262.3  &       &   1.7  &   T   &   e, f, H$\alpha$ only    \\
RP1753  &   RPJ 050619-690159   &   05 06 19.05 &   -69 01 59.4    &   4.0 &   285.1  &   16.7   &     &   T   &   e, s, H$\alpha$ only    \\
RP1189  &   RPJ 050622-704155   &   05 06 22.45 &   -70 41 55.4    &   6.7 &   261.3  &       &   0.9  &   T   &   c, p, ring  \\
\end{tabular}
 \end{table*}

\begin{table*}

\begin{tabular}{|l|l|c|c|c|c|c|c|c|l|}
&{\it (cont'd)}\\
\hline \hline
Cat. Ref &   IAU Name &   RA    &   DEC   &  H$\alpha$ Dia  &   V   &   V$_{err}$ &   V$_{err}$    &   P  &   Comments    \\
        &            &      J2000  &   J2000  &   arcsec  & helio    & em   & ccf    &     &     \\
\hline
RP1519  &   RPJ 051549-680445   &   05 15 49.79 &   -68 04 45.8    &   6.2 &   302.4  &       &   0.8  &   T   &   c, f, diffuse, H$\alpha$ only   \\
RP1440  &   RPJ 051555-684503   &   05 15 55.43 &   -68 45 03.1    &   4.3 &   292.2  &       &   0.5  &   T   &   irreg, diffuse, f, H$\alpha$ only   \\
RP1378  &   RPJ 051611-692036   &   05 16 11.03 &   -69 20 36.5    &   6.1 &   264.3  &       &   0.7  &   T   &   e, b, bp    \\
RP1240  &   RPJ 051638-701411   &   05 16 38.69 &   -70 14 11.4    &   4.0 &   234.0  &       &   2.2  &   T   &   e, f, H$\alpha$ only    \\
RP1415  &   RPJ 051642-685541   &   05 16 42.05 &   -68 55 41.3    &   7.0 &   249.0  &       &   1.4  &   T   &   c, s, b, halo   \\
\end{tabular}
 \end{table*}

\begin{table*}

\begin{tabular}{|l|l|c|c|c|c|c|c|c|l|}
&{\it (cont'd)}\\
\hline \hline
Cat. Ref &   IAU Name &   RA    &   DEC   &  H$\alpha$ Dia  &   V   &   V$_{err}$ &   V$_{err}$    &   P  &   Comments    \\
        &            &      J2000  &   J2000  &   arcsec  & helio    & em   &  ccf   &     &     \\
\hline
RP1093  &   RPJ 052231-664621   &   05 22 31.92 &   -66 46 21.6    &   3.5 &   351.4  &   44.1   &     &   L   &   poss c, half hidden \\
RP686   &   RPJ 052238-700420   &   05 22 38.93 &   -70 04 20.0    &   4.7 &   245.4  &   17.5   &     &   T   &   e, p, partly hidden \\
RP979   &   RPJ 052238-675508   &   05 22 38.96 &   -67 55 08.6    &   6.0 &   315.4  &   17.5   &     &   T   &   c, b, extended halo \\
RP757   &   RPJ 052243-693828   &   05 22 43.94 &   -69 38 28.1    &   8.7 &   254.6  &   29.4   &     &   T   &   c, b    \\
RP1113  &   RPJ 052249-664055   &   05 22 49.80 &   -66 40 55.5    &   8.1 &   323.0  &   21.7   &     &   P   &   e, VLE,  in \HII~region  \\
\end{tabular}
 \end{table*}

\begin{table*}

\begin{tabular}{|l|l|c|c|c|c|c|c|c|l|}
&{\it (cont'd)}\\
\hline \hline
Cat. Ref &   IAU Name &   RA    &   DEC   &  H$\alpha$ Dia  &   V   &   V$_{err}$ &   V$_{err}$    &   P  &   Comments    \\
        &            &      J2000  &   J2000  &   arcsec  & helio    & em   &   ccf  &     &     \\
\hline
RP1983  &   RPJ 05283.-695425   &   05 28 43.933    &   -69 54 25.3    &       &       &       &       &   P   &   RP695   \\
RP695   &   RPJ 052844-695422   &   05 28 44.10 &   -69 54 22.2    &   5.4 &   277.9  &       &   0.8  &   T   &   poss c, hidden  \\
RP1080  &   RPJ 052846-665440   &   05 28 46.97 &   -66 54 40.2    &   8.0 &   317.8  &       &   0.8  &   T   &   e, b, f diffuse halo    \\
RP2182  &   RPJ 05288.-690354   &   05 28 58.76 &   -69 03 54.9    &   10.1    &       &       &       &   P   &   e, b, in \HII~region \\
RP735   &   RPJ 052905-694657   &   05 29 05.06 &   -69 46 57.4    &   4.6 &   266.1  &       &   1  &   T   &   e, p    \\
\end{tabular}
 \end{table*}

\begin{table*}

\begin{tabular}{|l|l|c|c|c|c|c|c|c|l|}
&{\it (cont'd)}\\
\hline \hline
Cat. Ref &   IAU Name &   RA    &   DEC   &  H$\alpha$ Dia  &   V   &   V$_{err}$ &   V$_{err}$    &   P  &   Comments    \\
        &            &      J2000  &   J2000  &   arcsec  & helio    & em   &  ccf   &     &     \\
\hline
RP791   &   RPJ 053307-692946   &   05 33 07.62 &   -69 29 46.1    &   4.0 &   260.1  &   18.6   &     &   L   &   c, s, f, weak H$\alpha$ \\
RP1084  &   RPJ 053308-664806   &   05 33 08.73 &   -66 48 06.3    &   20.0    &   275.9  &      &   2.9  &   T   &   c arc, H$\alpha$ only, near b star  \\
RP548   &   RPJ 053308-711803   &   05 33 08.75 &   -71 18 03.3    &   4.7 &   225.7  &   26.7   &     &   T   &   e, s, p \\
RP908   &   RPJ 053323-683933   &   05 33 23.15 &   -68 39 33.8    &   3.8 &   334.2  &   21.1   &     &   L   &   c, p, s, stars N \& S   \\
RP604   &   RPJ 053328-712450   &   05 33 28.13 &   -71 24 50.7    &   4.7 &   199.2  &   20.3   &     &   T   &   c, f, diffuse, H$\alpha$ only   \\
\end{tabular}
 \end{table*}

\begin{table*}

\begin{tabular}{|l|l|c|c|c|c|c|c|c|l|}
&{\it (cont'd)}\\
\hline \hline
Cat. Ref &   IAU Name &   RA    &   DEC   &  H$\alpha$ Dia  &   V   &   V$_{err}$ &   V$_{err}$    &   P  &   Comments    \\
        &            &      J2000  &   J2000  &   arcsec  & helio    & em   &  ccf   &     &     \\
\hline
RP89    &   RPJ 053856-702121   &   05 38 56.41 &   -70 21 21.6    &   6.0 &   244.5  &   19.5   &     &   T   &   e, H$\alpha$ only   \\
RP246   &   RPJ 053857-693356   &   05 38 57.19 &   -69 33 56.5    &   7.0 &   260.4  &       &   1.3  &   P   &   c, in \HII~region    \\
RP219   &   RPJ 053902-693509   &   05 39 02.75 &   -69 35 09.3    &   3.0 &   266.1  &   33.3   &     &   P   &   c, p, in \HII~region \\
RP182   &   RPJ 053905-695045   &   05 39 05.12 &   -69 50 45.9    &   7.0 &   266.3  &       &   1.4  &   L   &   e, diffuse, H$\alpha$ only  \\
RP1951  &   RPJ 053906-711943   &   05 39 06.94 &   -71 19 43.5    &   6.0 &   270.4  &       &   1.8  &   L   &   irreg, diffuse, f, H$\alpha$ only   \\
\end{tabular}
 \end{table*}

\begin{table*}

\begin{tabular}{|l|l|c|c|c|c|c|c|c|l|}
&{\it (cont'd)}\\
\hline \hline
Cat. Ref &   IAU Name &   RA    &   DEC   &  H$\alpha$ Dia  &   V   &   V$_{err}$ &   V$_{err}$   &   P  &   Comments    \\
        &            &      J2000  &   J2000  &   arcsec  & helio    & em   &  ccf   &     &     \\
\hline
RP637   &   RPJ 054437-691947   &   05 44 37.27 &   -69 19 47.5    &   5.3 &   263.3  &       &   1.9  &   T   &   c, s, H$\alpha$ only    \\
RP189   &   RPJ 054448-693859   &   05 44 48.10 &   -69 38 59.8    &   10.0    &   261.0  &       &   0.8  &   P   &   e, in \HII~region    \\
RP252   &   RPJ 054500-691819   &   05 45 00.21 &   -69 18 19.2    &   5.5 &   263.4  &       &   3.5  &   P   &   c, f, VLE   \\
RP396   &   RPJ 054501-673034   &   05 45 01.64 &   -67 30 34.0    &   11.0    &   290.7  &   18.9   &     &   T   &   c, p diffuse    \\
RP263   &   RPJ 054511-691013   &   05 45 11.01 &   -69 10 13.7    &   3.8 &   261.2  &   23.6   &     &   P   &   c, s, VLE, in \HII   \\
\hline
\end{tabular}

\end{table*}

\label{section 8}


\section{Conclusions}

We present the positions of 460 new true, likely and possible PNe in
the Large Magellanic Cloud (LMC), together with accurate positions,
velocities, diameters and brief object comments. We have begun to
use this new population to explore the kinematics of the central bar
region of the LMC and find that the PNe population occupies a warped
disk with it's own kinematic line of nodes which is also perturbed
in a similar tangential direction but to a greater extent than the
\HI~gas. The measurement and comparison of PNe and \HI~indicates
that the innermost 2.6 deg bar region has non-circular motions. We
may therefore conclude that the area between the central main bar
and the larger LMC disk is slightly warped. The line of nodes is
somewhat twisted at the SE and NW edges. We see small evidence for a
warping of the line of nodes in the SW as found by Olsen \& Salyk
(2002). All the PNe and other objects within the RP survey are soon
to be available on the world-wide web including our H$\alpha$/SR
merged images and low resolution spectra.

\label{section 9}

\section*{Acknowledgments}

The authors wish to thank the AAO board for observing time on the
AAT and UKST. The authors also thank the European Southern
Observatory for observing time on the VLT, the South African
Astronomical Observatory and Australian National University along
with their telescope time allocation committees for supporting our
programme of follow-up spectroscopy. WR thanks Macquarie University,
Sydney, for travel grants. WR thanks Suzanne Reid for designing a
database to store the RP catalogue. WR thanks Rhys Morris and Fred
Dulwich for their kind assistance during the SAAO 1.9m observations.
The authors thank George Jacoby for his careful reading of the
paper.

\label{section 10}

\bsp

\label{lastpage}


\begin{thebibliography}{99}

\bibitem[\protect\citeauthoryear{}{}]{}
Barnes J.E., 1998, in Galaxies: Interactions and Induced Star
Formation, ed. R.C. Kennicutt, Jr., F. Schweizer \& J.E. Barnes (New
York: Springer), 275

\bibitem[\protect\citeauthoryear{Bland-Hawthorn, Shopbell \& Malin}{1993}]{Bland}
Bland-Hawthorn J., Shopbell P.L., Malin D.F., 1993, AJ, 106, 2154B

\bibitem[\protect\citeauthoryear{Bohannan \& Epps}{1974}]{Bohan}
Bohannan B.E., Epps H.W., 1974, A\&AS, 18, 47

\bibitem[\protect\citeauthoryear{Boroson}{1989}]{Boroson}
Boroson T.A. \& Liebert J., 1989, ApJ 339, 844

\bibitem[\protect\citeauthoryear{Chu}{1984}]{Chu}
Chu Y.H., Kwitter K.B., Kaler J.B., Jacoby G.H., 1984, PASP, 96, 598

\bibitem[\protect\citeauthoryear{Cohen}{1988}]{Cohen}
Cohen R.S., Dame T.M., Garay G., Montani J., Rubio M., Thaddeus P.,
1988, ApJ., 331, 95

\bibitem[\protect\citeauthoryear{}{}]{}
Dickey J.M., Mebold U., Marx M., Amy S., Haynes R.F., Wilson W.,
1994, A\&A, 289, 357

\bibitem[\protect\citeauthoryear{Dopita}{1991}]{Dopita}
Dopita, M.A., Ford H.C., Lawrence C.J., Webster B.L., 1985, ApJ.,
296, 390

\bibitem[\protect\citeauthoryear{Dopita}{1988}]{Dopita}
Dopita, M.A., Meatheringham, S.J., Webster, L.B., Ford, H.C., 1988,
ApJ., 327, 639



\bibitem[\protect\citeauthoryear{Einasto}{1988}]{Einasto}
Einasto J., Haud U., J$\hat{o}$eveer M., 1979, IAUS., 84, 231E

\bibitem[\protect\citeauthoryear{Feast}{1968}]{Feast}
Feast, M.W., 1968, MNRAS., 140, 345

\bibitem[\protect\citeauthoryear{Feast}{1984}]{f1}
Feast M.W., 1984, MNRAS, 211P, 51F

\bibitem[\protect\citeauthoryear{}{}]{}
Feitzinger J.V. 1980 pkdg. conf, 435F

\bibitem[\protect\citeauthoryear{}{}]{}
Feitzinger J.V. 1983 IAUS. 100, 241F

\bibitem[\protect\citeauthoryear{Fensen}{1985}]{f2}
Fesen R. A., Blair W. P., Kirshner R. P. 1985, ApJ., 292, 29F

\bibitem[\protect\citeauthoryear{Freeman}{1983}]{f3}
Freeman K.C., 1970, ApJ., 160, 811

\bibitem[\protect\citeauthoryear{Freeman}{1983}]{f4}
Freeman K,C., Illingworth G. and Oemler A. 1983, ApJ., 272, 488

\bibitem[\protect\citeauthoryear{}{}]{}
Gerin M., Combes F., Athanassoula E., 1990, A\&A, 230, 37

\bibitem[\protect\citeauthoryear{}{}]{}
Gooch R., 1996, Astronomical Data Analysis Software and Systems V,
A.S.P. Conference Series, Vol. 101, 1996, George H. Jacoby and
Jeannette Barnes, eds., p. 80.


\bibitem[\protect\citeauthoryear{}{}]{}
Hambly N. C., MacGillivray H. T., Read M. A., Tritton S. B., Thomson
E. B., Kelly B. D., Morgan D. H., Smith R. E., Driver S. P.,
Williamson J., and 4 coauthors, 2001, MNRAS, 326, 1279H

\bibitem[\protect\citeauthoryear{}{}]{}
Henize K.G., 1956, ApJS, 2, 315

\bibitem[\protect\citeauthoryear{}{}]{}
Jacoby G.H., 1980, AJ Suppl. Series, 42, 1

\bibitem[\protect\citeauthoryear{}{}]{}
Jacoby G.H., Ford H.C., 1983, ApJ, 266, 298

\bibitem[\protect\citeauthoryear{}{}]{}
Jacoby G.H., Walker A.R., Ciardullo R., 1990, ApJ, 365, 471J

\bibitem[\protect\citeauthoryear{}{}]{}
Jacoby G.H., 2006, in conf. proc. "Planetary Nebulae Beyond the
Milky Way" ed. L. Stanghellini, J. Walsh, N.G. Douglas
(Springer-Verlag) p.17

\bibitem[\protect\citeauthoryear{}{}]{}
Kaler J.B., Jacoby G.H., 1990, BAAS, 22R 1249K

\bibitem[\protect\citeauthoryear{}{}]{}
Kennicutt R.C., Bresolin F., French H., Martin P., 2000. ApJ, 537,
589

\bibitem[\protect\citeauthoryear{}{}]{}
Kim S., Staveley-Smith L., Dopita M.A., Freeman K.C., Sault R.J.,
Kesteven M.J., McConnell D., 1998, ApJ., 503, 674


\bibitem[\protect\citeauthoryear{}{}]{}
Klein U., Haynes R.F., Wielebinski R., Meinert D., 1993, A\&A 271,
402

\bibitem[\protect\citeauthoryear{}{}]{}
Kroupa P., Bastian U., 1997, New A., 2, 77

\bibitem[\protect\citeauthoryear{Kurtz}{1991}]{K1}
Kurtz M.J., Mink D.J., Wyatt W.F., Fabricant D.G., Torres G., Kriss
G.A., Tonry J.L., 1991, in Worrell D.M. et al., eds, ASP Conf. Ser.
Vol. 25, Astronomical Data Analysis Software and Systems I. Astron.
Soc. Pac., San Trancisco, p 432


\bibitem[\protect\citeauthoryear{}{}]{}
Kurtz M.J., Mink D.J., 1998, ASP, 110, 934


\bibitem[\protect\citeauthoryear{}{}]{}
Leisy P., Dennefeld M., Francois, P., 2000, ign. confE., 32L

\bibitem[\protect\citeauthoryear{}{}]{}
Lewis, I. J., Cannon, R. D., Taylor, K., Glazebrook, K., Bailey, J.
A., Baldry, I. K., Barton, J. R., Bridges, T. J., Dalton, G. B.,
Farrell, T. J., Gray P.M., Lankshear A., plus 11 authors, 2002,
MNRAS, 333, 279

\bibitem[\protect\citeauthoryear{}{}]{}
Lindsay E.M., 1963, Irish AJ, 6, 127

\bibitem[\protect\citeauthoryear{}{}]{}
Lindsay E.M., Mullan D.J., 1963, Irish AJ, 6, 51

\bibitem[\protect\citeauthoryear{}{}]{}
Luks T., Rohlfs K. 1992, A\&A, 263, 41

\bibitem[\protect\citeauthoryear{}{}]{}
Madore B.F., Freedman W.L., 1998, salg. conf. 263M

\bibitem[\protect\citeauthoryear{}{}]{}
Mathewson D.S., Ford V.L., 1984, in IAU Symposium 108, Structure and
Evolution of the Magellanic Clouds, ed. S. van den Bergh and K.S. de
Boer (Dordrecht:Reidel), p.125

\bibitem[\protect\citeauthoryear{}{}]{}
Mathewson, D.S., Ford V.L., Visvanathan N., 1986, ApJ., 301, 664

\bibitem[\protect\citeauthoryear{}{}]{}
McGee R.X. 1964, AuJPh. 17, 515M

\bibitem[\protect\citeauthoryear{}{}]{}
McGee R.X., Milton J.A. 1966, Aust. J. Phys., 19, 343

\bibitem[\protect\citeauthoryear{Meatheringham}{1988}]{m1}
Meatheringham S.J., Dopita M.A., Ford H.C. and Webster B.L. 1988,
ApJ., 327,651

\bibitem[\protect\citeauthoryear{Meatheringham}{1991}]{m2}
Meatheringham S.J., Dopita M.A. 1991, AJ suppl., 76, 1085

\bibitem[\protect\citeauthoryear{}{}]{}
Morgan D.H., Good A.R., 1992, A\&AS 92, 571

\bibitem[\protect\citeauthoryear{}{}]{}
Morgan D.H., 1994, A\&AS 103, 235

\bibitem[\protect\citeauthoryear{}{}]{}
Morgan, D.H., and Parker, Q.A. 1998, MNRAS., 296, 921

\bibitem[\protect\citeauthoryear{}{}]{}
Mould, J.R., Huchra J.P., Freedman W.L., Kennicutt R.C. Jr.,
Ferrarese L, Ford H.C., Gibson B.K., Graham J.A., Hughes S.M.G.,
Illingworth G.D., Kelson D.D., Macri L.M. and 5 coauthors, 2000,
ApJ, 529, 786

\bibitem[\protect\citeauthoryear{}{}]{}
Murai T., Fujimoto M., 1980, Pub. Astr. Soc. Japan, 32, 581

\bibitem[\protect\citeauthoryear{}{}]{}
Olsen K.A.G., Salyk C. 2002, AJ, 124, 20450

\bibitem[\protect\citeauthoryear{}{}]{}
Parker Q.A., Bland-Hawthorn J., 1998, PASA, 15, 33p

\bibitem[\protect\citeauthoryear{}{}]{}
Parker Q.A., Malin D., 1999, PASA, 16, 288P

\bibitem[\protect\citeauthoryear{}{}]{}
Parker Q.A., Phillipps S., Pierce M. J., Hartley M., Hambly N. C.,
Read M. A., MacGillivray H. T., Tritton S. B., Cass C. P., Cannon R.
D., Cohen M., Drew J.E. et al. 2005, MNRAS, 362, 689

\bibitem[\protect\citeauthoryear{}{}]{}
Pasquini L., Avila G., Blecha A., Cacciari C., Cayatte V., Colless
M., Damiani, F., de Propris R., Dekker H., di Marcantonio P.,
Farrell T., Gillingham P., Guinouard I., et al. 2002, Msngr., 110,
1P

\bibitem[\protect\citeauthoryear{Reid}{2006}]{r1}
Reid W.A. and Parker Q.A., 2006,
MNRAS, 365, 401R

\bibitem[\protect\citeauthoryear{}{}]{}
Rohlfs K., Kreitschmann J., Siegmann B.C., Feitzinger J.V. 1984,
Astr. Ap., 137, 343 (RKSF).

\bibitem[\protect\citeauthoryear{}{}]{}
Rudnick G., Rix H.W., 1998, ApJ, 116, 1163

\bibitem[\protect\citeauthoryear{}{}]{}
Sanduleak N., MacConnell D.J., Davis Philip A.G., 1978, PASP, 90,
621

\bibitem[\protect\citeauthoryear{}{}]{}
Sanduleak N., 1984, Structure and evolution of the Magellanic
Clouds, IAU Symp. 108, 231

\bibitem[\protect\citeauthoryear{}{}]{}
Smith M.G., Weedman D.W., 1972, ApJ, 177, 595S

\bibitem[\protect\citeauthoryear{}{}]{}
Swaters R.A., Schoenmakers R.H.M., Sancisis R., van Albada T.S.,
1999, MNRAS, 304, 330

\bibitem[\protect\citeauthoryear{}{}]{}
Tonry J.L., Davis M., 1979, AJ. 43, 393

\bibitem[\protect\citeauthoryear{}{}]{}
van der Marel R.P., Cioni M-R.L., 2001, AJ, 211, 1807V

\bibitem[\protect\citeauthoryear{}{}]{}
van der Marel R.P., 2001, ApJ, 122, 1827


\bibitem[\protect\citeauthoryear{}{}]{}
Vassiliadis E., Meatheringham S.J., Dopita M.A., 1992, ApJ., 394,
489

\bibitem[\protect\citeauthoryear{}{}]{}
Visvanathan N., 1985, ApJ, 288, 182

\bibitem[\protect\citeauthoryear{}{}]{}
Webster B.L., 1969, MNRAS, 143, 97

\bibitem[\protect\citeauthoryear{}{}]{}
Westerlund B.E., and Smith L.F., 1964, MNRAS, 127, 449

\bibitem[\protect\citeauthoryear{}{}]{}
Westerlund B.E. 1997, The Magellanic Clouds (Cambridge: Cambridge
Univ. Press)

\bibitem[\protect\citeauthoryear{}{}]{}
Zaritsky D., Rix H.W., 1997, ApJ, 477, 118

\bibitem[\protect\citeauthoryear{}{}]{}
Zhao H.S., Evans N.W., 2000, ApJ, 545, L35

\end{thebibliography}
\end{document}